\documentclass[12pt,preprint]{aastex} 
\newcommand{\lapprox}{$\stackrel {<}{_{\sim}}$}

\slugcomment{To appear in Ap.J.}

\begin{document}

\title{Pre-Main Sequence variables in the VMR-D : identification of
T Tauri-like accreting protostars through {\it Spitzer}-IRAC variability
}

\author{
   T.Giannini\altaffilmark{1},
   D.Lorenzetti\altaffilmark{1},
   D.Elia\altaffilmark{2,3},
   F.Strafella\altaffilmark{3},
   M.De Luca\altaffilmark{1,4},
   G.Fazio\altaffilmark{5},
   M.Marengo\altaffilmark{5},
   B.Nisini\altaffilmark{1},
           and
   H.A.Smith\altaffilmark{5}
}
%
\altaffiltext{1}{INAF - Osservatorio Astronomico di Roma, via
Frascati 33, I-00040  Monte Porzio, Italy, giannini,
dloren,nisini,deluca@oa-roma.inaf.it}
\altaffiltext{2}{Universidade de Lisboa, Facultade de Ciencias,
Centro de Astronomia e Astrofisica da Universidade de Lisboa,
Observatorio Astronomico de Lisboa, Tapada da Ajuda1349-018,
Lisboa, Portugal,eliad@oal.ul.pt}
\altaffiltext{3}{Dipartimento di Fisica, Univ.del Salento, CP 193,
I-73100 Lecce, Italy, eliad,francesco.strafella@le.infn.it}
\altaffiltext{4}{LERMA-LRA, UMR 8112, CNRS, Observatoire de Paris and
Ecole Normale Supérieure, 24 Rue Lhomond, 75231 Paris, France, massimo.de.luca@lra.ens.fr}
\altaffiltext{5}{Harvard-Smithsonian Center for Astrophysics,
Cambridge, MA 02138, mmarengo@cfa.harvard.edu,
hsmith@cfa.harvard.edu}
%


\begin{abstract}
We present a study of the infrared variability of young stellar objects by means
of two {\it Spitzer}-IRAC images of the 
Vela Molecular Cloud D (VMR-D) obtained in observations separated 
in time by about six months.  
By using the same space-born IR instrumentation, this study 
eliminates all the unwanted
effects due to differences in sensitivity, confusion,
saturation, calibration, and filter band-passes, issues that are usually
unavoidable when comparing catalogs obtained from different instruments. The 
VMR-D map covers about 1.5 deg$^2$ of a site
where star formation is actively ongoing. We are interested in accreting pre-main sequence variables whose luminosity 
variations are due to intermittent events of disk accretion (i.e.
active T Tauri stars and EXor type objects). The variable objects have been selected from
a catalog of more than 170,000 sources detected at a S/N ratio $\ge$5.  
We then searched the sample of variables for ones whose photometric properties such as IR excess, color-magnitude relationships, and spectral energy distribution, are as 
close as possible to those of known EXor's. Indeed, these latter are monitored in a more
systematic way than T Tauri stars and the mechanisms that regulate the observed phenomenology
are exactly the same. Hence the modalities of the EXor behaviour is adopted as driving 
criterium for selecting variables in general. We ultimately selected 19 {\it bona fide}
candidates that constitute a well-defined sample of new variable
targets for further investigation (monitoring, spectroscopy).
Out of these, 10 sources present a {\it Spitzer} MIPS 24\,$\mu$m counterpart, and 
have been classified as 3 Class~I, 5 flat spectrum and 2 Class~II objects, while the spectral 
energy distribution of the other 9 sources is compatible with evolutionary phases older than 
Class~I. 
This is consistent with what is known about the small sample of known EXor's,
whose properties 
have driven the present selection and suggests that the accretion flaring or EXor stage might come
as a ClassI/ClassII transition.
We present also new prescriptions that can be useful in future 
searches for accretion variables in large IR databases.
\end{abstract}




%

\section{Introduction}

Young stellar objects (YSO's) from low to intermediate mass (0.5-8
M$_{\sun}$) accumulate the majority of their final mass during
the so-called main accretion phase lasting about 10$^5$ years.
After this period they continue to accrete, although at lower
rates. According to a commonly accepted view, material that
falls onto the circumstellar disk crosses the viscous
part of the disk itself and finally jumps onto the stellar surface
through the magnetic interconnection lines (Shu et al. 1994).
These accretion phenomena are observationally signaled by
intermittent outbursts (usually detected in the optical and/or in
the near-IR bands) due to the sudden increase of the mass
accretion rate by orders-of-magnitude (Hartmann \& Kenyon 1985).
These events are quite common during the Pre-Main Sequence life (T
Tauri and Herbig Ae/Be stars) and, when some recurrent flaring events
are recognizable in their light curves, they are
called FUor or EXor outbursts depending on their intensity and duration 
(Herbig 1989; Hartmann \& Kenyon 1996). 
FUor's are rarer, and experience particurlarly intense outbursts (5-6 mag) followed by long
periods (tens of years) during which the object remains bright.
Conversely, typical EXor variability (Herbig 2007, 2008;
Lorenzetti et al. 2006, 2007, 2009, hereinafter L07, L09) and the variability of 
accreting T Tauri stars is characterized by less intense (3-4 mag)
and more short-lived outbursts (from months to a year) superposed on 
longer periods (few years) of quiescence. Historically, all the known
accreting T Tauri stars, FUor's, and EXor's were first identified in the visual bands; as a
consequence, being quite un-extincted, they have been usually
associated with the last phases of pre-main sequence evolution.
However, there is no physical reason that prevents the associated phenomena
from occurring as well during more embedded, and consequently
earlier, phases. In fact, an increasing number of similar objects
has been identified and studied (e.g. Hodapp et al. 1996, 1999;
Reipurth \& Aspin 2004; K\'{o}sp\'{a}l et al. 2007;
Sicilia-Aguilar et al. 2008). Obviously there are many kinds of variable objects besides 
accretion and eruptive variables: cataclysmic, pulsating (Mira), and 
eclipsing variables, for example. Hence, although accreting ones are expected to
prevail in star formation regions, great care has
to be taken evaluating possible contamination effects.


Sofar, the low-mass disk accretion variables are extensively studied only in nearby star
forming regions (at distances less than 300 pc), but now Spitzer sensitivity has enlarged
the accessible volume, thus allowing to investigate that population at larger distances, as
well. Such a chance is relevant for investigating the interplay between low-, intermediate- 
and even high-mass stars formation processes occurring in the same region.

We recently used the InfraRed Array Camera (IRAC,
Fazio et al. 2004) and the Multi-band Imaging Photometer for
Spitzer (MIPS, Rieke et al. 2004) on board of {\it Spitzer Space
Telescope} (Werner et al. 2004) to survey 1.5 deg$^2$ of the Vela Molecular
Cloud-D (Giannini et al. 2007), a star forming region in the
southern sky on which we have accumulated a large variety of
photometric (Massi et al. 1999, 2000, 2003, De Luca et al. 2007)
and spectroscopic (Giannini et al. 2005; Lorenzetti et al. 2002)
data, along with detailed maps of the gas (Elia et al. 2007) and
dust (Massi et al. 2007) emission. The total 
integration time requested for the IRAC map was 
obtained by co-adding two individual maps, each integrating for
half of the total time and separated by about one semester in time. This observational mode additionally provided us with 
two separate sets of data ideally suited for investigating 
the mid-infrared variability (over 6 months) of our 
sample of young objects. While this multiple epochs technique has been commonly  
used to acquire Spitzer/IRAC maps of star forming regions, to allow  
easy removal of transients and artifacts, these datasets have been  
seldom analyzed to study the variability of the sources.
Given the features that characterize accreting T Tauri stars, EXor's and FUor's, 
the relatively short period between the two IRAC data sets does not seem suitable 
for a systematic search for FUor's, that usually remain stable for decades. 
Hence in the following discussion, when speaking of accreting variables we will refer
to the class of active T Tauri stars or even to EXor's, given their similarities
in the accretion modalities.

The advantages of our approach are many: ({\it i})
the time elapsed between the two maps (6 months) is well suited 
for sampling the most common accretion events; ({\it ii}) because IRAC wavelengths 
are less affected by extinction than optical ones, IRAC allows us to enlarge 
the effective volume under investigation compared to optical observations, while 
preserving the needed sensitivity;
({\it iii}) the IRAC bands (3.6, 4.5, 5.8 and 8.0 $\mu$m) cover
the spectral range where disks emit the largest part of their
energy; ({\it iv}) a comparison between two sets of data obtained
with the same instrumentation removes the confusion introduced by 
instrumental differences of sensitivity, beam confusion, saturation levels, 
calibration, and/or filter band-passes.

Our primary goal is to select valid candidates of variable accreting objects in
VMR-D, a star forming region in which such a search has never 
before been done.  Once completed, we expect that photometric 
and spectroscopic monitoring will be in order to confirm and analyze
their nature.  Secondly, we want+ to use our sample to provide useful 
distinguishing prescriptions for finding similar sources in other star 
forming regions, and thus to increase the number of known objects in a systematic way.
In this respect we remind that our
analysis can not account for the complete population of YSO's in
accretion, since they spend the major part of their lifetime in
quiescence or in a steady state.
A thorough analysis in that sense, designed to select YSO's within the large 
{\it Spitzer} sample of midplane objects by reducing contamination from 
spurious sources, has been recently presented by Robitaille et al.
(2008).

This paper follows the following structure: after having presented
our observations (Sect.2) and defined the variable objects
(Sect.3), we discuss our results (Sect.4) with emphasis on the
source evolutionary stage and location inside the cloud. Finally, we present our conclusions (Sect.5).

\section{Observations}

In this paper we make use of a map of approximately 1.5 deg$^2$
of the Vela Molecular Cloud-D that we obtained with {\it Spitzer}-IRAC
Cycle-3 GTO (PID: 30335; PI: G.Fazio) at 3.6, 4.5, 5.8 and 8.0
$\mu$m. The area chosen for mapping is the same we have 
investigated at other wavelengths (see Sect.1). Observations were 
carried out in two separate AORs on 2007 February 21 UT and
2007 July 4 UT (AOR Keys: 17606656 and 17606912, respectively) by
adopting half-array cross scan offsets.  The cross scans also enabled us to obtain much more reliable point source photometry in crowded regions.

\begin{figure}
 \centering
   \includegraphics [width=15 cm] {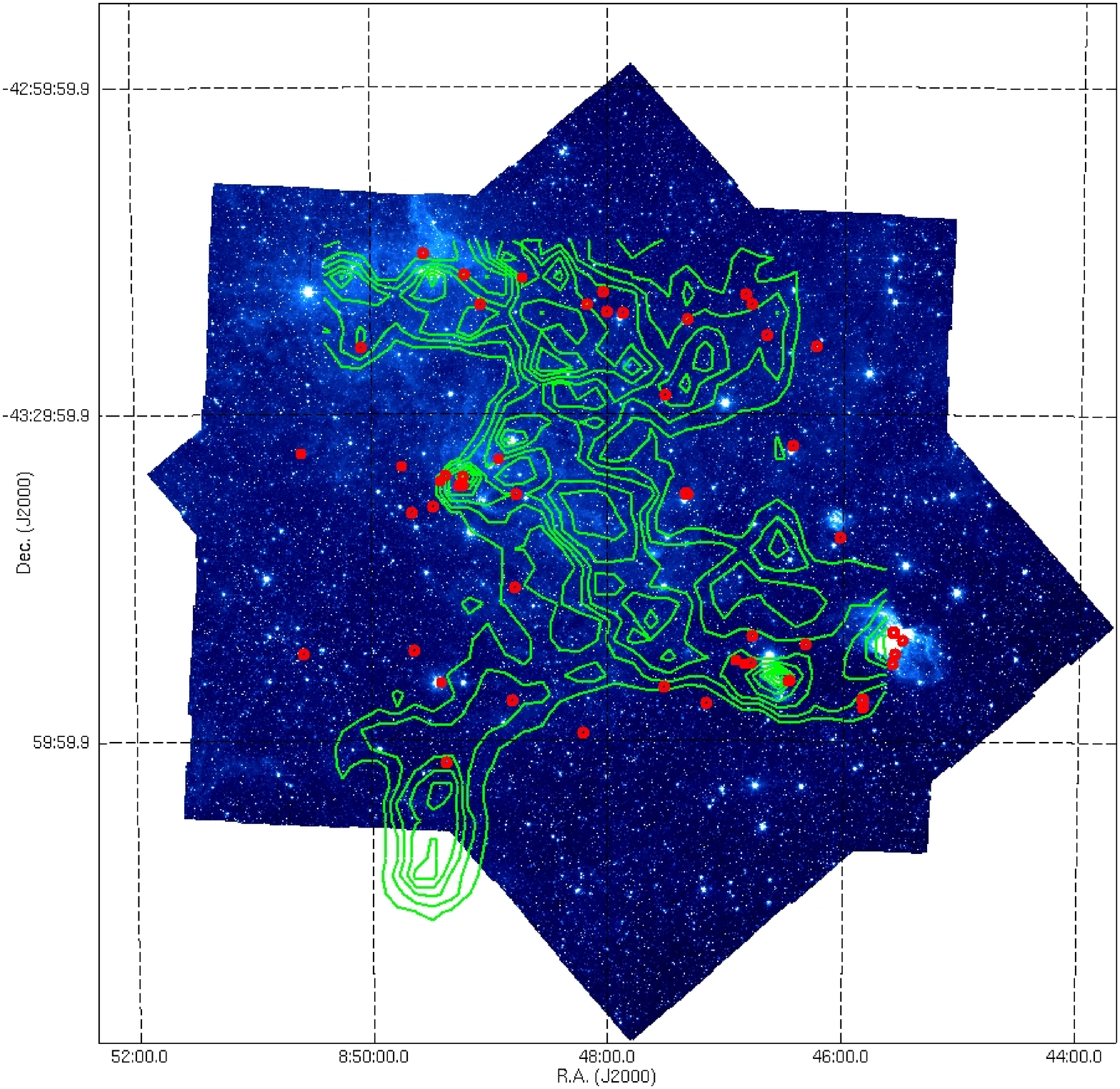}
   \caption{IRAC maps at 3.6\,$\mu$m obtained in two different epochs. Due to the satellite 
   orientation, they are
   rotated in a different way. The common part is
   clearly recognizable. The contours of our CO map (Elia et al. 2007) are superposed (in green). 
   The 5.8\,$\mu$m channel is acquired simultaneously, therefore it presents the same 
   superposition. The location of the selected variables is shown with red dots.
   \label{map1:fig}}
\end{figure}

\begin{figure}
 \centering
   \includegraphics [width=15 cm] {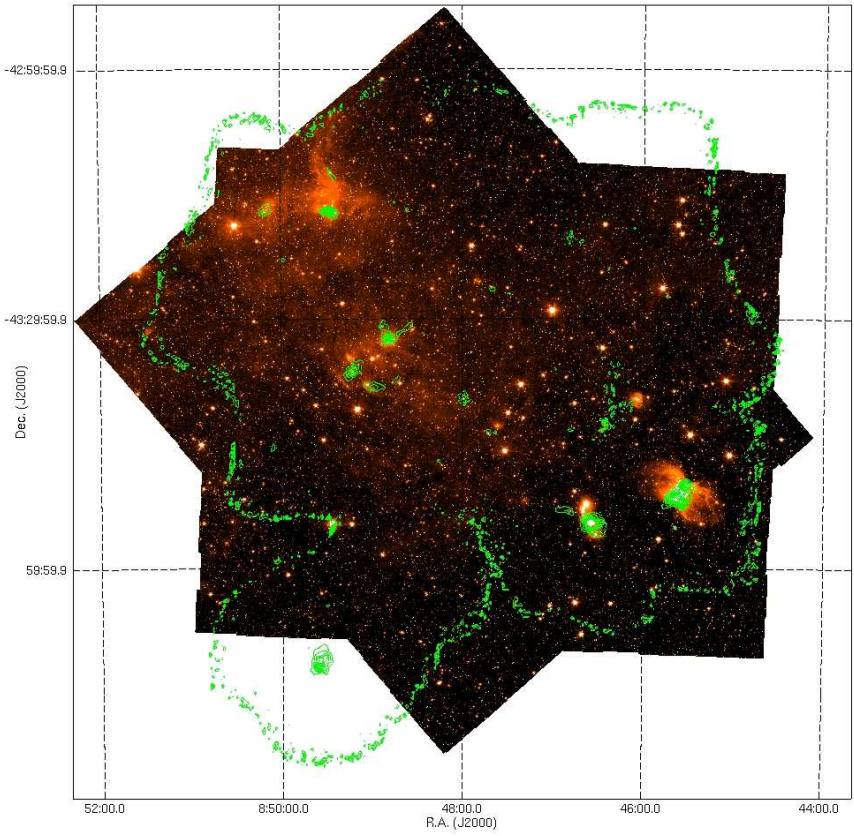}
   \caption{As in Figure \ref{map1:fig}, but for the 4.5\,$\mu$m
   channel (acquired simultaneously to the 8.0\,$\mu$m one).
   Notice that the common part between the two epochs is slightly different from
   that of channels 3.6 and 5.8\,$\mu$m. Here, for completeness, our
   dust continuum map at 1.2 mm (Massi et al. 2007) is superposed (in green).
   \label{map2:fig}}
\end{figure}

A set of 256 frames, each 10.4~s integrated, was used, resulting in a total
integration time of 20.8~s per pointing. In regions where bright sources were expected we obtained a few frames in High Dynamical Range (HDR) mode to obtain unsaturated fluxes. 
The IRAC data were reduced using the IRACproc package (Schuster et  
al. 2006), to obtain a single flux calibrated mosaic combining all the  
individual exposures in each epoch (separately), on a pixel grid of  
0.8627"/pix. IRACproc is based on the Spitzer Science Center mosaic  
software MOPEX and provides enhanced outlier (cosmic rays) rejection.  
The IRAC point-source photometry was done using DAOPHOT package (Stetson 1987). Image
mosaicing and all the details relative to both the PSF photometry
extraction and the construction of a catalog containing more than
170,000 IRAC sources are presented in a dedicated paper (Strafella
et al. 2009). Here we note that the spatial resolution of {\it Spitzer} 
in the IRAC bands is of the order of 2$\arcsec$. 
Because of the different satellite
orientations in the two epochs, the maps do not
coincide precisely, as shown in Figures~\ref{map1:fig} and \ref{map2:fig};
moreover, band 1 (3.6\,$\mu$m) is acquired simultaneously with band
3 (5.8\,$\mu$m), while band 2 (4.5\,$\mu$m) simultaneously with band 4
(8.0\,$\mu$m); also these couples of maps are not exactly coincident
although to a much
lesser extent. As a result, in the following we will refer only to
those sources detected in the common portion (between both
different epochs and different bands), in order to have, for any
considered source and for any considered band, two observations
available for comparison.

\section{Identification of variable objects}

Since we are interested in selecting variable YSO's, the two
independent sub-catalogues were searched using the
following procedures. Firstly, the variable objects within a given
band were defined with these selection criteria:

\begin{itemize}
\item[-] detection in the considered band at a S/N level $\geq$
5 in both epochs, separately
\item[-] ratio between the flux variation and its error,
(F1-F2)/$\sigma$(F1 - F2) $\geq$ 5
\item[-] object brighter than a given threshold in both periods; 
namely $\leq$ 16 mag (band 1), $\leq$ 15.5 mag (band 2), $\leq$ 13.5
mag (band 3), $\leq$ 12.5 mag (band 4)
\end{itemize}

\begin{figure}
   \includegraphics [width=15 cm] {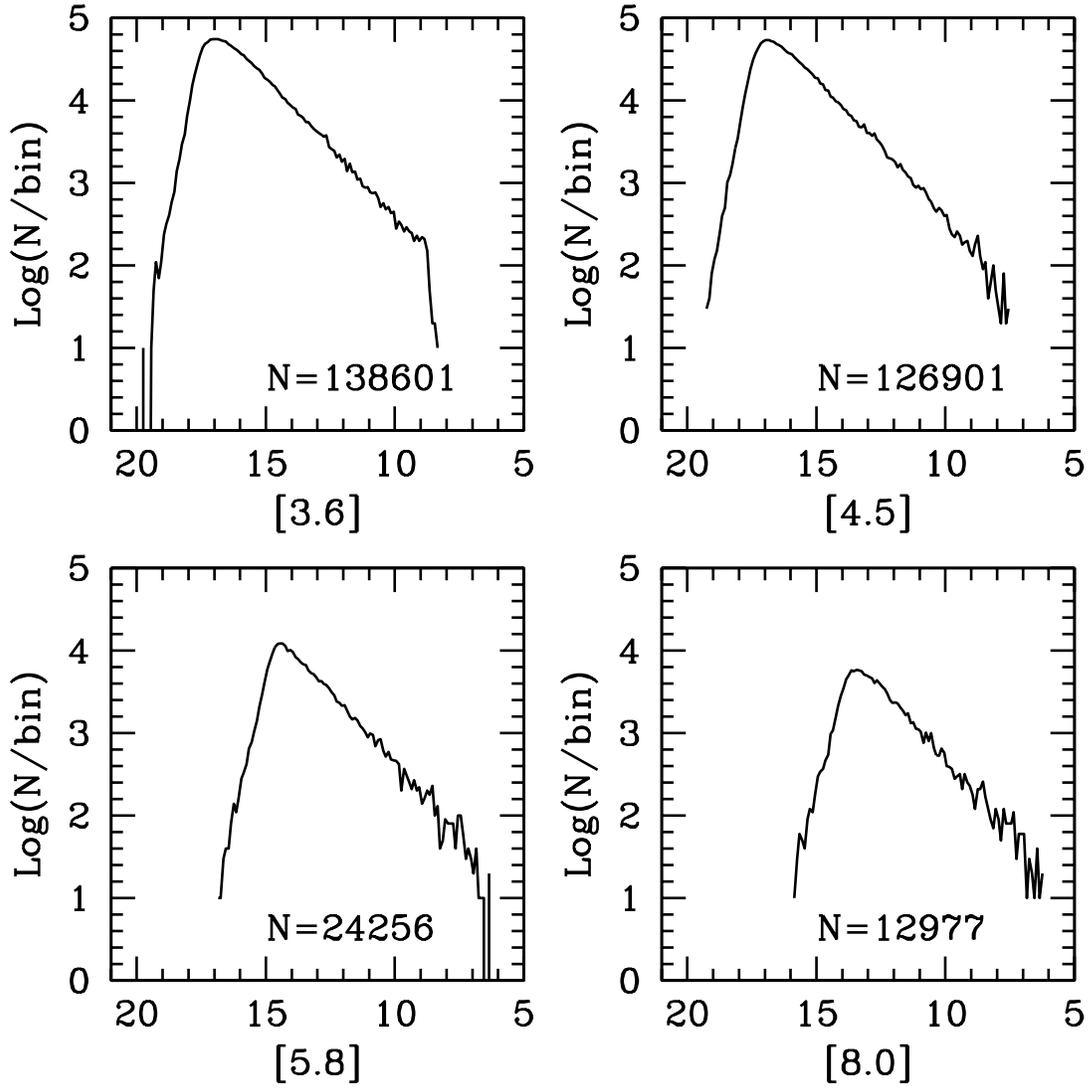}
   \caption{Magnitude distribution of all the IRAC sources in VMR-D as detected
   in the first epoch (see text). The same
   distribution relative to the second epoch does not present any
   significant difference.
   \label{complP1:fig}}
\end{figure}

The first criterium picks up objects associated with real detections, minimizing any 
contamination by unwanted effects such as spikes and artifacts. The second one allows 
us to select genuine variations well above the photometric
errors. The third one is dictated by the completeness limits of
our sub-catalogues (see Strafella et al., 2009, and
Figure~\ref{complP1:fig}). Moreover, since our sample of selected variables
will plausibly need near-IR spectroscopy to be definitely validated, from the start we 
decided only to select sources bright enough to be accessible with the current 
spectroscopic instrumentation.

Statistical results on the detection rate of variables are
given in Table~\ref{stat:tab}, where we tabulate, for each band, the number of the 
sources detected in the first epoch (column 1, this number does not significantly change in 
the second epoch), the sources above the magnitude threshold (column 2) and the 
variables that satisfy all the three above criteria (column 3). Noticeably, the number of
variables in the first two bands are more than double than those in band 3 and 4.

Negative (positive) magnitude variations correspond to declining (rising) 
events between the first and the second epoch. 
In all the four bands, the number of declining sources is 
significantly larger than that of rising ones (see columns 5 and 6 of Table~\ref{stat:tab}). Ã
This result suggests that the declining time is typically longer than the rising 
one roughly by a factor 1.3.  

\begin{figure}
 \centering
   \includegraphics [width=15 cm] {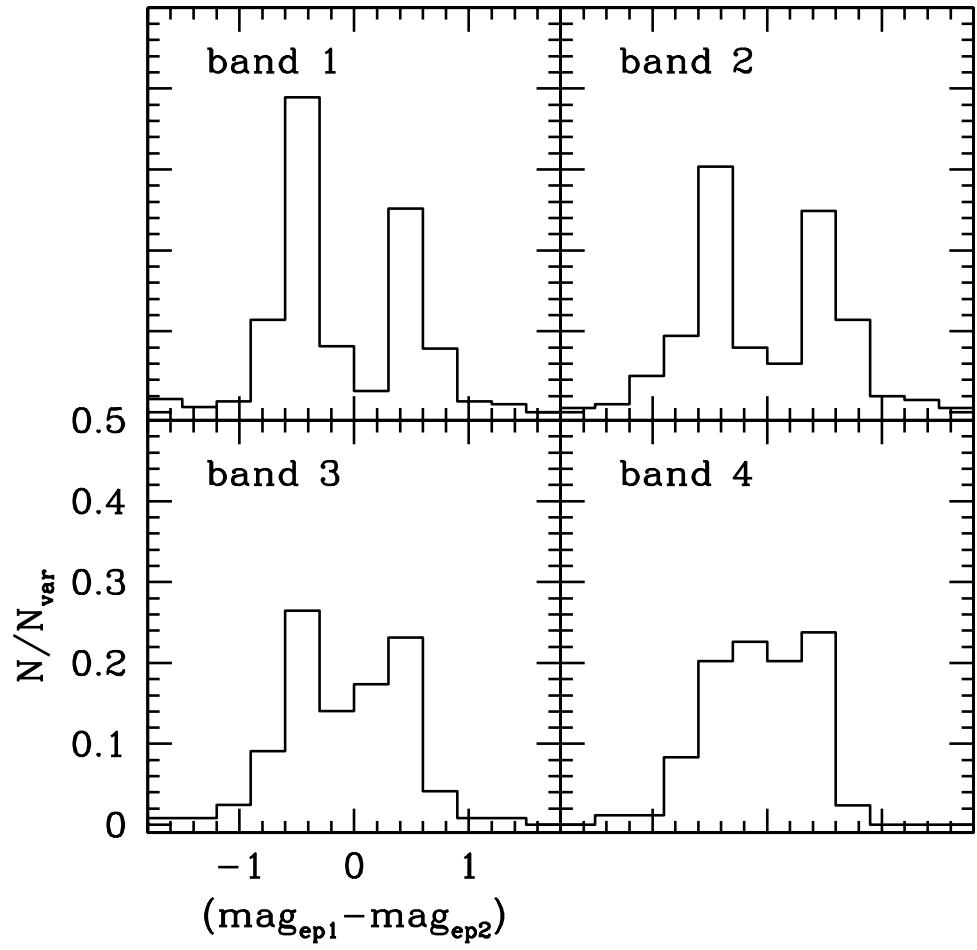}
   \caption{Distribution of the variable IRAC sources as a function of the magnitude
   variation.
   \label{histo:fig}}
\end{figure}

In Figure~\ref{histo:fig} the distribution of variable
sources (normalized to the total number of variables in that band)
as a function of the magnitude variation is given. 
Each of the four distributions shows two peaks around a variation of $\sim$ 0.4-0.6 magnitudes, roughly symmetrical around zero.
This effect is clearly a consequence of our second selection criterium
that validates only those variations significantly larger than their own errors. 

We note that the gap in the distribution tends to be filled at the longer
wavelength bands, both because of the increase of the level of the minimum around zero, 
and because of the progressive decrease in height of the two peaks. This indicates that small percentage intensity variations occur preferentially at the longer wavelengths. A similar behavior has already been
identified in the set of known eruptive young objects (although at different wavelengths, 
i.e.\,in the near-infrared bands): they are never observed to vary only in a single spectral 
band, and their percentage fluctuations usually decrease in amplitude with wavelength (L07). 

The above considerations lead us to pick up among the objects
that passed our criteria, those simultaneously
variable in both bands 1 and 2, remaining in this way with 53 sources.
These have been examined one by one
looking for their location in the original
3.6\,$\mu$m mosaic and ruling out those sources located in the near
proximity of very bright objects. In principle, PSF photometry
should be less sensitive to contamination from nearby
contributions, however artifacts associated to bright IRAC objects
present different patterns in the two different epochs and this
occurrence may mimic some spurious variability. As a result of
such an inspection six further sources have been dropped out, hence
our final sample is constituted by a total of 47 objects, which represent 
the 0.22\% of the detections (above the magnitude threshold)
in both bands.
These 47 variable objects will be considered in the following and
a catalog of these sources is given in Table~\ref{mag:tab}. Here,
IRAC magnitudes (one line for each epoch) are given along with
complementary (and {\it not} contemporary) photometry of the 2MASS (JHK)
and MIPS (24\,$\mu$m) counterparts (see below).

In order to verify whether or not the selected sources are intrinsically
variable or perhaps appear as such because of some fluctuations of the
local sky or background between the two epochs (for example inside the IR
clusters, where the very localized diffuse emission may be relevant), 
we have plotted the variations (in band 1 and 2) of all the IRAC sources within
1$\arcmin$ from the selected variable. By doing so, we have
temporary relaxed our second criterium (otherwise no other
variable could be found in the neighborhood). 
We did not find any source whose flux fluctuation was comparable with that
of the selected objects. This demonstrates that our sample is 
constituted by genuine variables, whose variation cannot be ascribed 
to any other phenomenum not related with the source itself.

\section{Results and Discussion}

\subsection{Location of the variable sources}

Some or all of the 47 selected sources could in principle be 
foreground or background objects not related to the young population of VMR-D. In Figure~\ref{map1:fig} the
location of these sources, with respect to the distribution of the
molecular gas, is shown. Most of them are seen to be distributed
near the IR clusters, or near the peaks of the warm gas.
We take this as confirmation that we are really tracing a young population of variable sources
in VMR-D. In the 
Table~\ref{summary:tab} (2nd column) a flag is given for each object indicating whether it lies outside the CO contours (O),
inside (I), or exactly within the peaks (P). We use this
information to build up a {\it decision Table} which
will help to clarify the nature of the sources and provide a summary 
of all the evidences from our analysis. The aim of this approach is to 
investigate to what
extent our sample presents characteristics similar to those of known EXor's, 
as derived 
from our previous studies (L07, L09) and here adopted as typical drivers for selecting 
disk accreting objects.


\subsection{Contamination}
\begin{figure}
 \centering
  \includegraphics [width=15 cm] {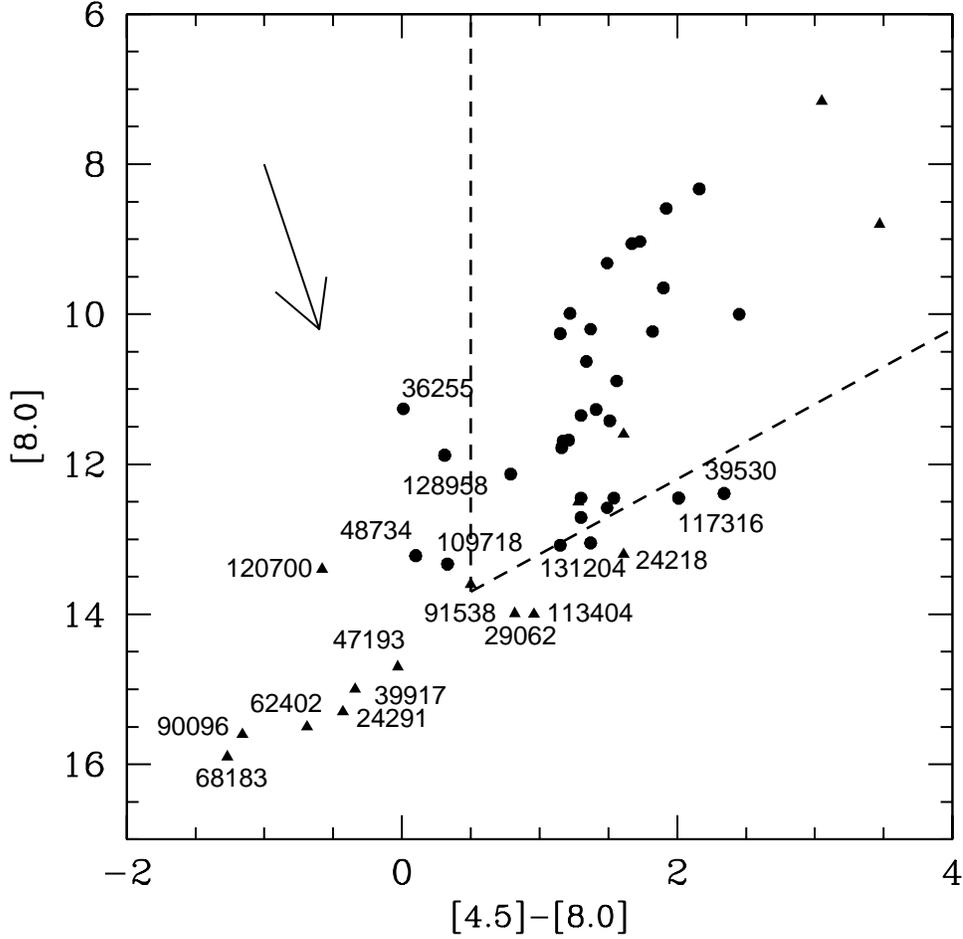}
   \caption{Color-magnitude diagram of the 47 selected variables. In this diagram YSO's are located
   between the lines [4.5]-[8.0]$>$0.5 and [8.0]$<$14-([4.5]-[8.0]), i.e. the open triangle at the
   top-right. Dots represent objects detected
in both considered bands; triangles indicate the upper limits at 8.0\,$\mu$m: therefore
they should imagined to point toward the bottom-left corner of the figure. We have labeled the 
sources which lie outside the region occupied by YSO's, also flagged as extragalactic (XGAL) or
photospheres (PHT) in Table~\ref{summary:tab}. The arrow shows the effect of the 
extinction for A$_V$ = 50 mag.
   \label{colmag1:fig}}
\end{figure}

In addition to the morphological analysis above, 
we have applied more quantitative tools to separate the YSO's from
the remainder of the sources in the map.
Main sequence stars and background galaxies can easily be recognized in the 
[8.0] vs.[4.5]-[8.0] diagram (e.g. Harvey et al. 2006), reported in Figure \ref{colmag1:fig}.
Here the 47 variables are indicated with dots if observed in both of the two bands or 
with triangles if they have remained undetected at 8.0\,$\mu$m, in which case we have 
considered the 3\,$\sigma$ upper limit. 

In this diagram, the 
YSO's occupy the open triangle on the top-right side, while
the photospheres are located to the left of the vertical line (at [4.5]-[8.0] = 0.5) 
and the galaxies (whose {\it locus} has been established by the SWIRE ELAIS-N1 
extragalactic survey, Rowan-Robinson et al. 2004) on the bottom-right 
side (e.g. J{\o}rgensen et al. 2006, Porras et al. 2007).
Noticeably, the majority of the sources detected in the two bands lie in 
the YSO's region; on the contrary, just a few 8.0\,$\mu$m upper limits are compatible with it. 
A significant fraction of sources fall just below the oblique dashed line: these could be either faint YSO's or external galaxies. In any case, since we cannot ascertain their nature in a definite way, we have labeled them 
as possible extragalactic (XGAL) sources in Table\,\ref{summary:tab}, column\,3. 
Similarly, sources with  [4.5]-[8.0] $<$ 0.5 have been flagged as photospheres (PHT).

A harder class of objects to identify is the asymptotic giant branch (AGB) stars, that are Long Period Variables
(LPVs) of Mira semi-regular, or irregular type, and which represent an important
fraction of the population of variables sources in the plane of the Galaxy.
As shown by Marengo et al. (2008), AGB's are sources with IR excesses easily
detectable by IRAC. However, in the IRAC bandpasses they have numerous
molecular absorption features whose strengths and transition
frequencies depend on both their chemistry (essentially the
atmospheric C/O ratio) and mass loss rates. Such features can strongly
affect their IRAC colors, nominally expected to be those of reddened
photospheres, and make them to appear very similar to those of YSO's.
The timescale of AGB (Mira) variability is roughly comparable to
that of EXor's, and as a result the time elapsed between our two epochs is not 
helpful in separating AGB stars from YSO's.

The spatial distribution of our 47 sources by association with the
molecular cores reduces the contamination by AGB stars, which are 
not expected to cluster in star forming regions; nevertheless,
some additional attempts deserve to be done
to estimate how the AGB contamination affects our sample.
To this end, we have constructed 3 two-color diagrams ([3.6]-[4.5] versus [5.8]-[8.0], [4.5]-[5.8] versus [5.8]-[8.0]
and [3.6-4.5] versus [3.6]-[8.0]), marking on them the AGB {\it loci} as 
derived in Marengo et al. (2008). Then, we flagged as potential
AGB's all the sources that simultaneously fall inside the 3 AGB's {\it loci}
(or are in the close proximity to the contours) defined for each plot. 

Very recently, Robitaille et al. (2008) have carefully investigated 
the issue of
separating AGB's and YSO's, and provide specific criteria for
characterizing each class with IRAC and MIPS 24\,$\mu$m
photometry. One of these is that sources with
[8.0]-[24] $<$ 2.5 are candidate AGB stars, while those with 
[8.0]-[24] $>$ 2.5 are classified as YSO's. This means that
sources with no MIPS 24\,$\mu$m detection cannot be separated by their criteria, 
but most likely are dominated by YSO's. By applying their criteria to our
sample, only three sources (\#57549, 85679 and 154688) can be identified as AGB candidates. They reassuringly already belong to the sample we have 
flagged as potential AGB's on the basis of the IRAC color-color diagrams.
In Table~\ref{summary:tab} (column\,3) the candidate AGB's are labeled.

Finally, a considerable number of M-type dwarfs are expected in the foreground of VMR-D. The possible
contamination from this class of objects will be discussed further on (Sect.\,4.4).

\begin{figure}
 \centering
   \includegraphics [width=15 cm] {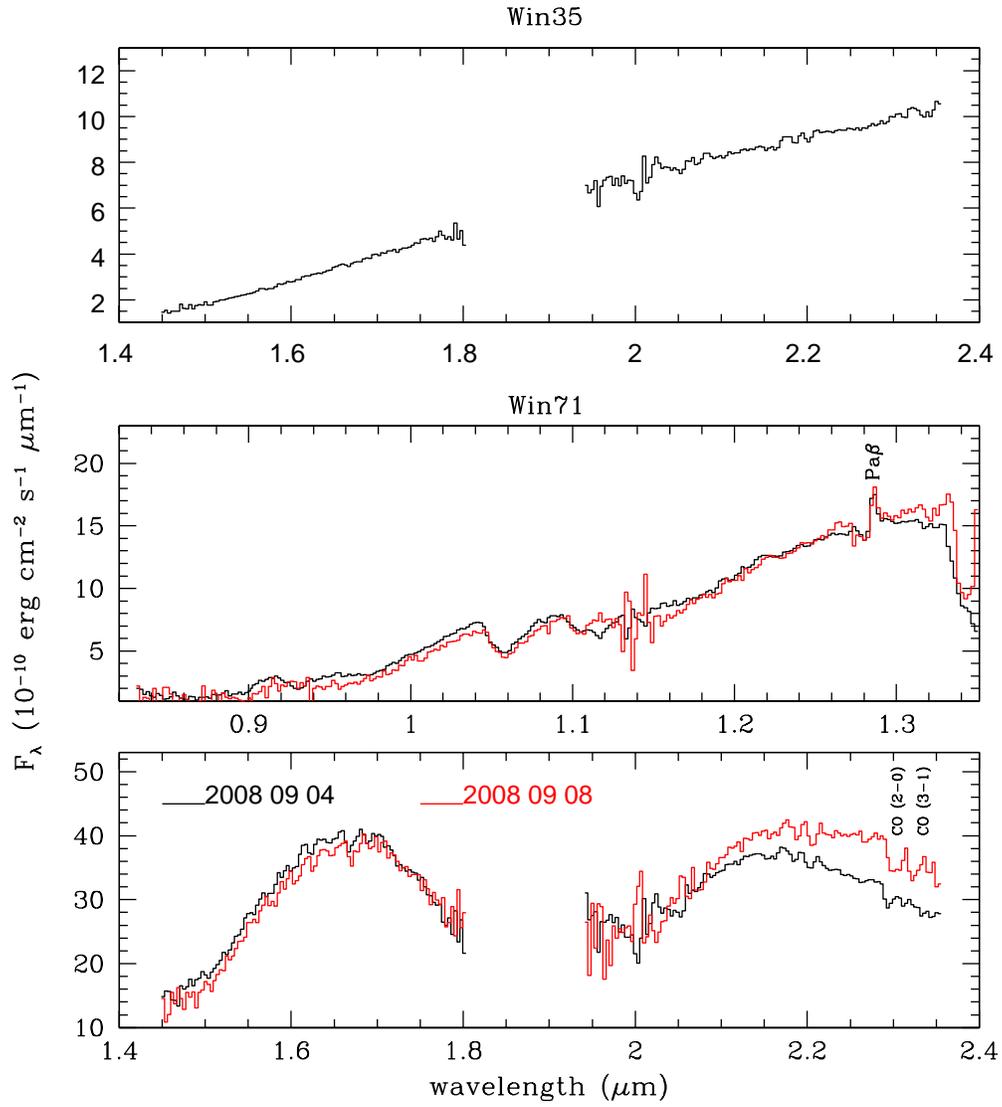}
   \caption{Near IR spectra of the sources win 35 (top) and win 71 (bottom).
   \label{wins:fig}}
\end{figure}

The ultimate test capable of proving the nature of the selected
sources is obviously spectroscopic. Indeed, by taking a
low resolution ($\mathcal{R}$ $\sim$ 200-300) near-IR spectrum
(1.0-2.5\,$\mu$m) of any individual source, we would be able to
determine if it is an emission line object (as all the T Tauri stars are, and all
the EXor's appear
to be - L09), a protostar with an increasing 
continuum with wavelengths, or an
AGB star characterized by deep absorption bands due to water
vapor at 1.4 and 1.9\,$\mu$m (and sometimes even at 2.5\,$\mu$m)
superposed on an increasing continuum. As an example, we show in 
Figure~\ref{wins:fig} the spectra of two bright sources belonging to a 
sample of protostellar candidates selected from IRAC/MIPS data
in the Serpens star forming region (Winston et al. 2007). Both these sources
\#35 and 71, here called win35 and win71 for simplicity)
were observed during a spectral survey we are carrying out at our 1m IR
telescope (at Campo Imperatore - Italy). As is apparent, win35 
behaves as a typical protostellar source, while win71 presents
unequivocally (and confirmed in two repeated spectra) the deep
H$_2$O absorption bands more typical of an AGB star (the possibility that
this source is a protostar of late spectral type can be disregarded because
of its brightness). The same conclusions have indeed reached by Winston et al. (2009),
who have recently undertaken a near-infrared spectroscopic survey of the YSO's in Serpens.
These examples serve to illustrate ({\it i}) how the IR colors alone, without any further
evaluation, can lead to possible misidentifications, and ({\it
ii}) that near-IR spectroscopy is an effective means of refining the 
identifications based on IRAC/MIPS colors alone.

\subsection{IRAC colors}

\begin{figure}
 \centering
   \includegraphics [width=15 cm] {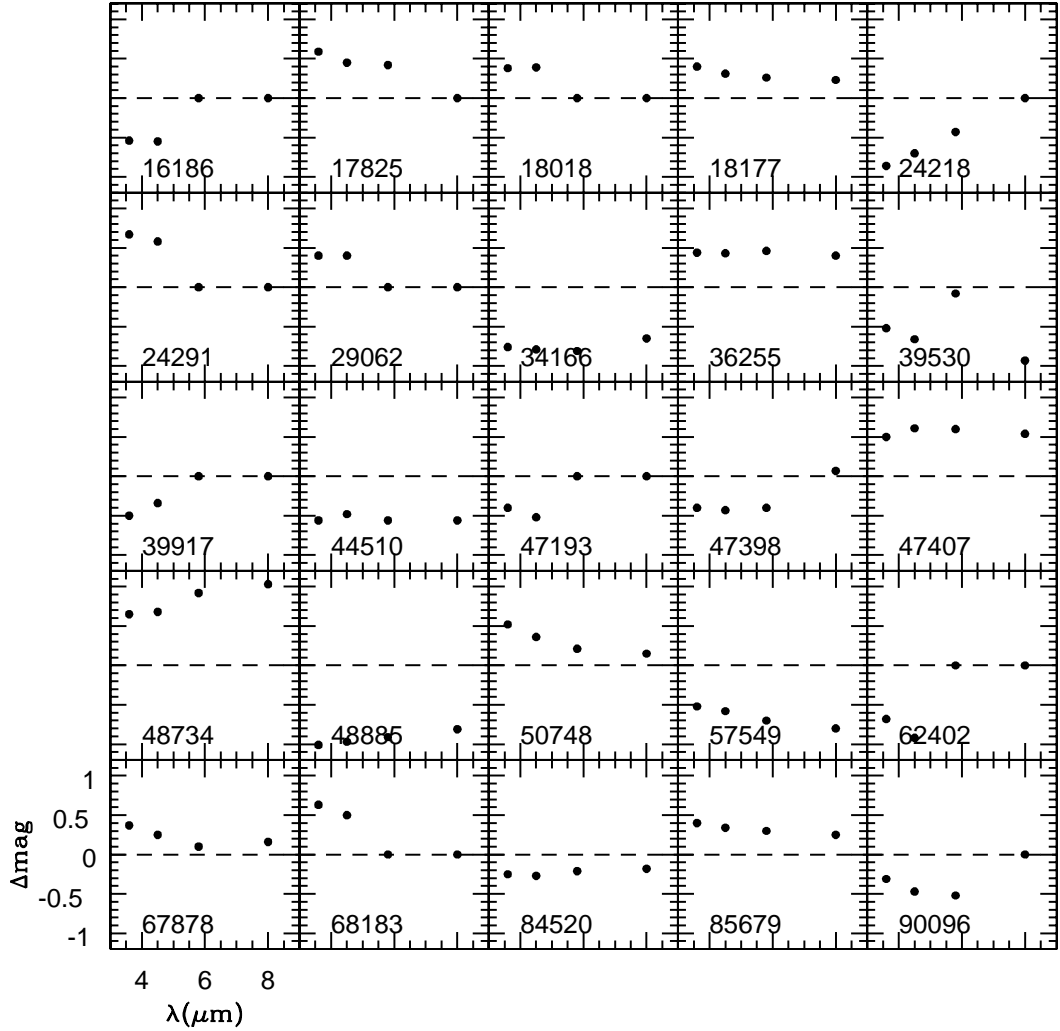}
   \caption{a). Magnitude variation of the 47 sources in the two epochs,
   $\Delta$mag = [mag(epoch1)-mag(epoch2)], as a function of the observed wavelength.
\label{deltamag1:fig}}
\end{figure}

The EXor's monitored so far in the near-IR (1-2.5\,$\mu$m) present two
main photometric characteristics (L07): {\it (i)} as most intrinsic variables, they become
bluer during the outburst and redder while they fade;
{\it (ii)} the amplitude of the variations decrease with increasing
the wavelength. These conclusions are based on observations of well-defined variability events, 
namely, those where the full amplitude of the variation is  $\Delta$mag $\ga$ 1mag,  and not to 
marginal fluctuations usually associated with randomly sampled events.  During the outburst phase
a peak at a relatively high temperature becomes more and more
evident (blueing), while marginal fluctuations can be ascribed to
colder contributions typical of a disk temperature stratification.
In the present sample, $\Delta$mag $\ga$ 1 variations are
quite rare (see Figure~\ref{histo:fig}).  Nevertheless, with the
above-mentioned caveat in mind, it is useful to search our sample
for some similarities (or differences) with known
EXor's, since here the lack of a systematic monitoring is
compensated by approaching a large and unbiased sample in a
statistical sense.  To that end, we present in Figures~\ref{deltamag1:fig}a and 
\ref{deltamag1:fig}b the
magnitude difference [mag(epoch1)-mag(epoch2)] as a function of the
wavelength, with variations measured between the first and the second epochs, 
while 
Figures~\ref{colmag2:fig}a and \ref{colmag2:fig}b depict the
[3.6]-[4.5] vs.\,[4.5] color-magnitude plots of the selected sources. 

From the data shown in these Figures some conclusions can be
derived:

{\it (i)} from Figures~\ref{deltamag1:fig}a and \ref{deltamag1:fig}b
we see that in 21 cases (representing $\sim$45\% of the total
sample) the object's amplitude decreases with increasing wavelength, as is typical of 
the monitored EXor's; 17 cases ($\sim$35\%) show comparable
variability at all wavelengths; while for the remaining 9 sources
(less than 20\%) the variation appears to increase with wavelength. Noticeably, 
just one of these latter objects is classified as YSO by means of the 
color-color and color-magnitude diagrams. 
These results are summarized in column\,4 of Table~\ref{summary:tab} with the following 
codes: D(ecreasing), C(onstant), I(ncreasing). They confirm that, for the majority of 
sources (38 out of 47), IRAC variability is consistent with that
displayed by EXor's;

{\it (ii)} irrespective of the source brightness (from 15 to 10
mag at 4.5\,$\mu$m, see Figures~\ref{colmag2:fig}a and
\ref{colmag2:fig}b) the amplitude of the [3.6]-[4.5] color variation
is substantially similar (\lapprox 0.2 mag) for sources 
whose color lies in the range 0.0-1.0 mag (with only 7 exceptions of redder color);

{\it (iii)} as mentioned above, intrinsic variables in the near-IR tend
to show redder and redder colors while becoming fainter, although
this conclusion can only be univocally applied to sources with large
fluctuations. The opposite (i.e. the colors become bluer), happens when the source
becomes brighter. With reference to Figures~\ref{colmag2:fig}a and
\ref{colmag2:fig}b, both these variation modalities translate into a
negative slope of the line connecting the two data points. 
This information is also listed in column\,5 of Table~\ref{summary:tab},
coded as -, 0, + to indicate a negative, null or positive slope 
of this line. The majority of sources (37 out of 47) are fully compatible with known EXor
behavior, i.e are coded as - or 0.

\subsection{Spectral Energy Distributions}


We obtained the Spectral Energy Distributions (SED's) for the 47 selected variables by 
identifying counterparts at both shorter and longer wavelengths when possible. 
To identify possible counterparts in the JHK
bands, we used the Two Micron All Sky Survey (2MASS) catalog (Cutri et al.
2003), looking for matches within a radius of 2$\arcsec$.
We identified 21 sources with a valid 2MASS detection, and the
corresponding JHK magnitudes are given in Table~\ref{mag:tab}. We
also searched the literature (Massi et al. 2000) for JHK images at a better
sensitivity (magnitude limit is K $\sim$ 18) than 2MASS, but,
unfortunately, none of them overlaps the regions with the selected variables.
In the view of a future spectroscopic followup, we have also searched for optical counterparts: 
the positive detections are listed in a footnote of Table~\ref{mag:tab}.
To search for longer wavelength counterparts, we used the IRAS Point Source Catalog,  the
Midcourse Space Experiment (MSX) catalog (Price et al. 2001) and
our MIPS catalog (Giannini et al. 2007). No
association was found to any of our 47 sources in the former two, to within a radius of 20$\arcsec$, 
while 24\,$\mu$m MIPS counterparts were found in only 13 cases using a search radius of 6$\arcsec$
(also listed in Table~\ref{mag:tab}). Only
two of our sources (namely \# 121029 and 124521) have a 70\,$\mu$m counterpart in a search radius 
of 20$\arcsec$; however, the multiplicity of the IRAC sources 
in the MIPS 70\,$\mu$m beam implies that these associations are not unique, and we do not consider 
them in the following analysis.
Statistically, 50\% of the IRAC selected variables possess a
near-IR counterpart, 30\% have a 24\,$\mu$m one, and about 20\% have both.
 
Figures~\ref{sedA:fig}a and
\ref{sedA:fig}b show the SED's of the 47 variable sources, from 2MASS wavelengths to 
MIPS\,24\,$\mu$m 
(when available). The IRAC fluxes from both epochs are shown 
in different colors; black dots represent the complementary
photometry from different periods. Open squares indicate IRAC and MIPS 3\,$\sigma$ upper limits. 
For comparison, a median stellar photosphere in the
spectral range K5-M5 is also plotted in each panel (Hern\'{a}ndez et
al. 2007), normalized to the flux corresponding to the shortest infrared wavelength available, namely J or
3.6\,$\mu$m (this latter for the sources without a 2MASS counterpart). The last two panels show the
SED's of UZ Tau E and V1647 Ori: the first one (Hartmann et al.
2005) is a well known and quite unextincted nearby EXor, while the
second is a more embedded (and maybe more massive) candidate
observed with {\it Spitzer}/2MASS during an outbursting phase
(Muzerolle et al. 2005). These distributions, that appear to be fully consistent 
with that expected for accretion disks (e.g. D'Alessio et
al. 1999), can be used as a working template for comparing the SED's of our
selected variables, and for identifying among them the most
likely acccreting YSO's candidates. 
By examining the SED's of our variables, we can identify 
some objects as a late type photosphere; others
show an excess (more or less pronounced) at increasing
wavelengths, that is typical of a temperature
stratification due to the presence of an evolved circumstellar disk or envelope. 
We define, as a quantitative 
indicator, the ratio $\mathcal{E}$ between the observed SED and
the underlying median K5-M5 photosphere, both integrated from J to
8.0\,$\mu$m. For UZ Tau E and V1647 Ori we obtain $\mathcal{E}$ =
2.3 and 26.8, respectively. Therefore,  
we conservatively suggest 
that accreting YSO candidates are those objects with $\mathcal{E}$ $>$ 1.
The result of such a criterium is seen in
Table~\ref{summary:tab} (column 6), and can be summarized by saying
that more than 50\% of the 47 selected variables present a SED
compatible with an accretion disk. We notice that none of the objects with $\mathcal{E}$ $>$ 1
simultaneoulsy presents IRAC fluxes compatible with a photosphere and a MIPS excess: this implies that
possible binary systems composed by a variable photospheric object and a colder companion do not 
contaminate our sample.  

The shapes of the SED's also suggest that a non-negligible fraction of the 
luminosity is emitted at wavelengths longer than those probed with our observations.  
Consequently, we cannot give a reliable estimate of the bolometric luminosity. 
The absolute values of the luminosities in the IRAC range (L$_{IRAC}$, last column of
Table~\ref{summary:tab}) are, on average,
remarkably low (from thousandths to tenths of one solar luminosity)
when compared with the IRAC estimated luminosities of known eruptive variables (7.9 and 0.4 L$_{\sun}$
for V1647 Ori and UZ Tau, respectively). 
Given the high sensitivity of IRAC, this
circumstance is not surprising, since, having observed VMR-D just twice without performing a
systematic monitoring, 
we sample primarily the numerous, low luminosity end of the stellar distribution in the cloud.
The luminosities obtained in both epochs differ from each other by between 15 and 
50 \% (with a few larger exceptions). 
Such variations are consistent with $\dot{M}$ fluctuations
within a factor 2-4, exactly the values expected over these time-scales (L09). 

As anticipated in Sect.\,4.2, several M-type red-dwarf stars displaying emission lines (dMe) are expected in the
foreground of VMR-D, given its distance at about 700 pc. These dMe stars present both fluxes
comparable to those of our variables and flaring activity due to enhanced coronal emission.
However, their SED's are fitted by a purely photospheric spectrum up to 24\,$\mu$m and no evidence
for IR excess has been found (Riaz et al. 2006). Since a definite presence of IR excess is
required by our selection, our final sample of variables is not affected by dMe contamination.
We are however able to detect them. To roughly estimate a lower limit of (K-M) flaring
photospheres that lie on our investigated area, we notice that 10 objects (i.e. about 20\% of
the total) have been disregarded as accreting candidates, but display SED's typical to a
photospheric spectrum (see Figure~\ref{sedA:fig} and Table~\ref{summary:tab} for their
$\mathcal{E}$ value).

\subsection{Analogies with known eruptive variables}

Having now accumulated in Table~\ref{summary:tab} a range of 
independent markers for EXor behavior, or more simply of active T Tauri stars, 
we can now scrutinize these indicators in a systematic way in an attempt to identify the 
variables that behave as the young accretors do: namely, those that
simultaneously present all the characteristics summarized
in the last line of Table~\ref{summary:tab} and discussed in the
previous Sections.  After analyzing in this way each of the sources in
Table~\ref{summary:tab}, we conclude that 19 (boldfaced) sources out of 47
manifest the same five flags we argue are typical of EXors: P(I),
D(C), YSO, -(0), and $>$1.0. Hence, we identify these 19 as the accreting protostars candidates
emerging from the present study. In particular, 6 of them have the
same flags strictly (P, D, YSO, -, $>$1.0). Out of the remaining 
29 sources, another 10 show 4 flags 
in accordance with the template, while 19 sources
have three or fewer identifying flags.

From an evolutionary point of view, protostars can be classified 
according to the value of the spectral index computed from 2 to 10\,$\mu$m 
(Greene et al. 1994); however the classification scheme 
does not change substantially if computed up to fluxes of 20-25\,$\mu$m 
(e.g. Rebull et al. 2007). Among the final 19 objects, 10 are detected at 24\,$\mu$m.  
Of these, 3 are Class~I ($\#$ 17825, 44510, 50748), 5 are flat spectrum ($\#$ 67878, 84520, 107546, 110128, 125801)
and 2 are Class II sources ($\#$ 131555, 133791). This indicates that the accreting flaring stage
may occur earlier than the Class II phase, at least in the framework of an 
evolutionary scenario. For comparison, L07 have classified 2 EXor's 
as Class I sources (NY Ori and PV Cep), 3 as flat spectrum sources
(XZ Tau, V1143 Ori and EX Lup) and 4 as Class II (UZ Tau, VY Tau,
DR Tau and V1118 Ori). 
For the remaining 9 sources, 
we computed the slope of the SED up to the longest wavelength available
(typically 8\,$\mu$m) and consistent with the 24\,$\mu$m upper limit. 
In this way we find that 3 and 6 sources are compatible with (or older than) flat and Class~II
sources, respectively.

Our demography of the VMR-D cloud (Strafella et al. 2009) provides a total of
487 YSO's made up of Class~I (62), flat spectrum (92) and
Class~II (333) objects.  It is remarkable that in addition there are 
171 Class~III sources associated to the VMR-D. 
Because variables that were in quiescence during the epochs of our monitoring 
are missed by the present selection, we cannot reasonably attempt to estimate the 
duration of the intense disk accretion phase with respect to the 
entire PMS lifetime. We can, however, conclude that: {\it i}) by adopting our empirical
approach, about 50\% of the selected variables are
potential active T Tauri stars or EXor candidates deserving of further study (mainly IR
monitoring and spectroscopy);  {\it ii}) a value
of 4\% is a lower limit on the percentage of accreting variables with respect
to the total protostars (from Class~I to Class~II sources) in VMR-D; {\it iii})
an important fraction of these 19 candidates is constituted by objects
with Class I and flat spectra, a trend {\it opposite} to the young
stellar population in VMR-D which, on the contrary, is dominated
by Class II and III sources. This circumstance might stem from
our requirement that our candidates are requested to be associated with 
CO peaks or filaments: this results in 
finding more embedded and younger objects. Moreover, the need for 
a detectable excess ($\mathcal{E}$ flag) goes in the same
direction.  

\section{Concluding Remarks}

We present a catalog of objects identified as probable PMS variables,
obtained by comparing two IRAC images of the star forming region VMR-D separated by six months. By analyzing the results, we reach the following conclusions:

\begin{enumerate}
\item The same selection criteria were applied to the data-set of both epochs
aiming to find real detections (by removing spikes and artifacts) with genuine variations (well above the photometric errors). Recorded flux fluctuations span from 10\% to more than 100\%.
\item Variability in the first two IRAC bands (3.6 and 4.5\,$\mu$m) identifies more than 
twice as many objects than does variability in bands 3 and 4 (5.8 and 8.0\,$\mu$m).
\item 47 variable objects have been identified (out of a total of 170,000 sources).  
We estimate the possible contamination of the sample by considering extraneous galactic 
or extragalactic sources by using the well-defined properties of active stars (or EXor type) 
with recurrent disk accretion phenomena as templates.
\item Spectral Energy Distributions were constructed from near- to mid-IR for all the 47 sources.  
IRAC luminosities are remarkably low (from thousandths to tenths of one solar luminosity), 
but this is not surprising since, having observed the region just twice without performing a
systematic monitoring, we likely sample the low luminosity end of
the stellar distribution.
\item 19 sources from our full sample of 47 have all the same properties that characterize 
known EXor objects. They are potential accretion flaring or EXor candidates and  deserve further 
studies (mainly IR monitoring and spectroscopy) to better characterize their nature. In this latter
respect we note that the brightness of the selected variables is well compatible with 
the sensitivity of the current infrared spectroscopic instrumentation.
\item A significant number (i.e.\,8 sources) of the final 19 candidates are recognizable as
Class I and flat spectrum sources (with 2 Class II objects).
This suggests that the accretion flaring or EXor stage might come as a Class I/Class II transition.

\item  New prescriptions are derived from our analysis that can facilitate identifying  
accretion variables in large IR database.
\end{enumerate}

\begin{acknowledgements}
The authors would like to thank Arkady A.Arkharov and Valeri
M.Larionov for providing them with the near IR spectra of the
sources win35 and win70, taken at Campo Imperatore (Italy).
This work is based on observations made with the Spitzer Space Telescope, which is operated by the Jet Propulsion Laboratory,
California Institute of Technology under NASA contract
1407. Support for the IRAC instrument was provided by NASA
under contract number 1256790 issued by JPL.
\end{acknowledgements}

\addtocounter{figure}{-1}
\begin{figure}
 \centering
   \includegraphics [width=15 cm] {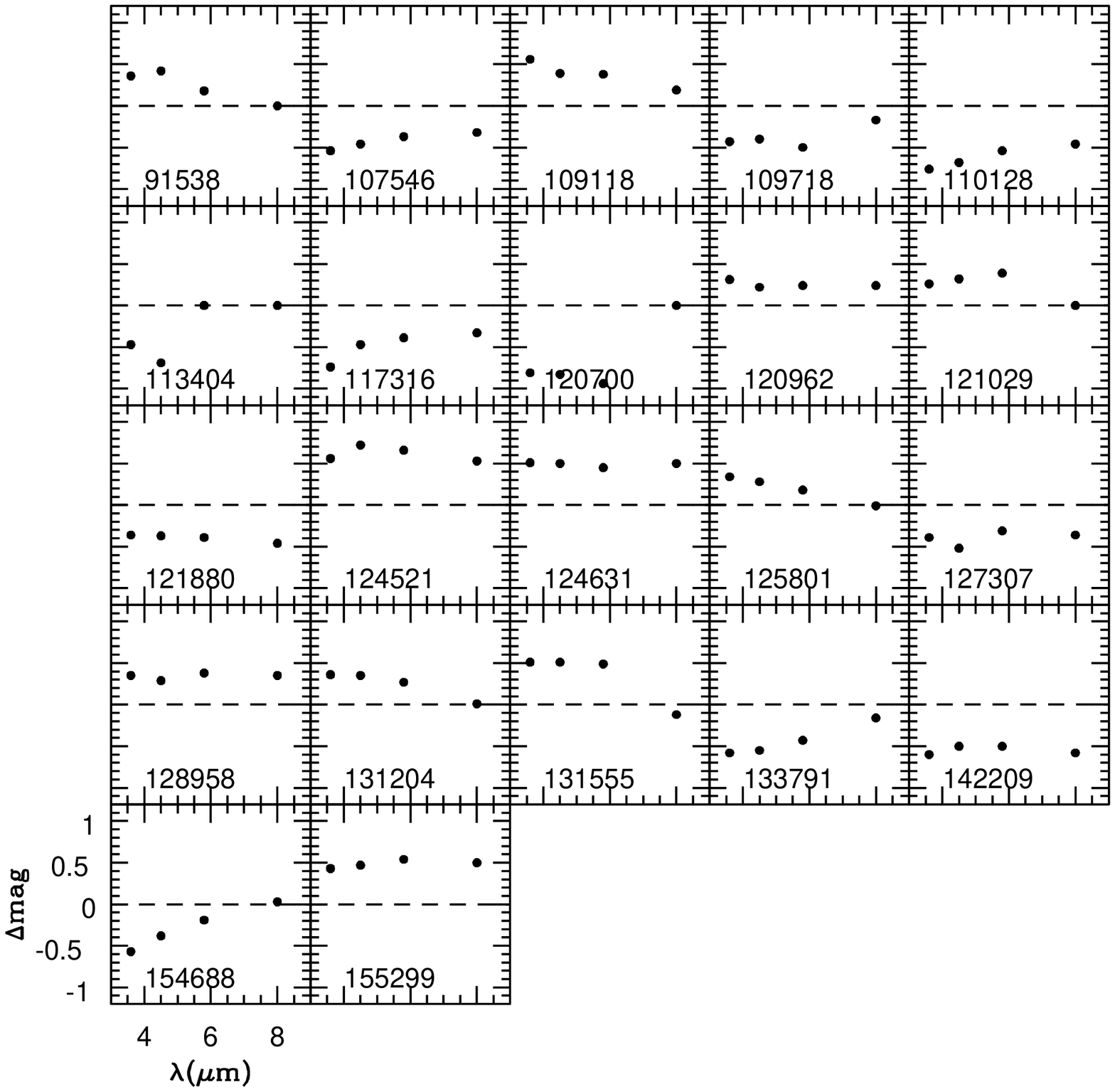}
   \caption{b). As Figure~\ref{deltamag1:fig}a, continued.
}
\end{figure}

\begin{figure}
 \centering
   \includegraphics [width=15 cm] {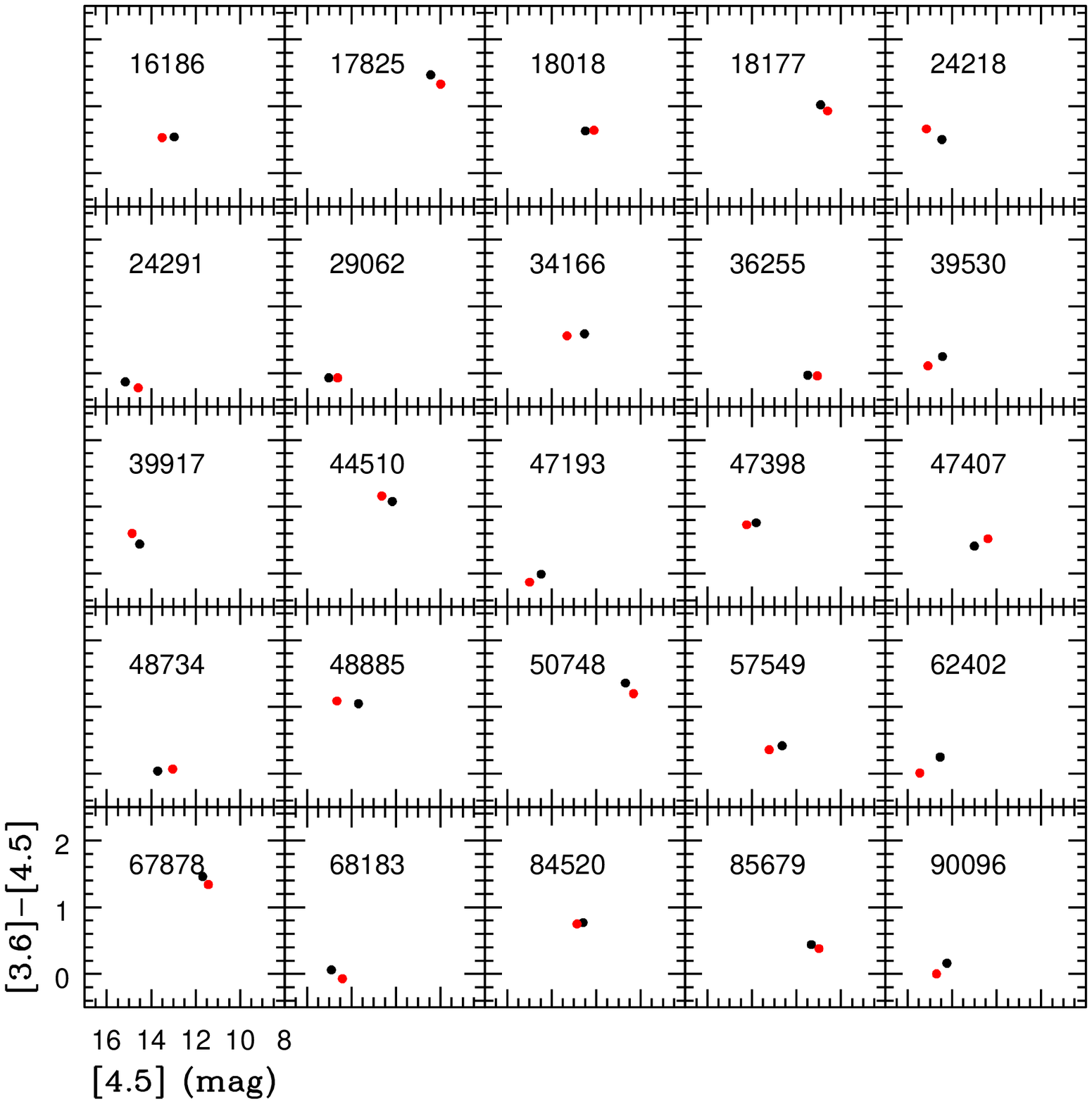}
   \caption{a). Observed color [3.6]-[4.5] vs. [4.5] magnitude of the 47 selected sources in both epochs.
   Black (red) dot refers to the first (second) epoch.
   \label{colmag2:fig}}
\end{figure}

\addtocounter{figure}{-1}
\begin{figure}
 \centering
   \includegraphics [width=15 cm] {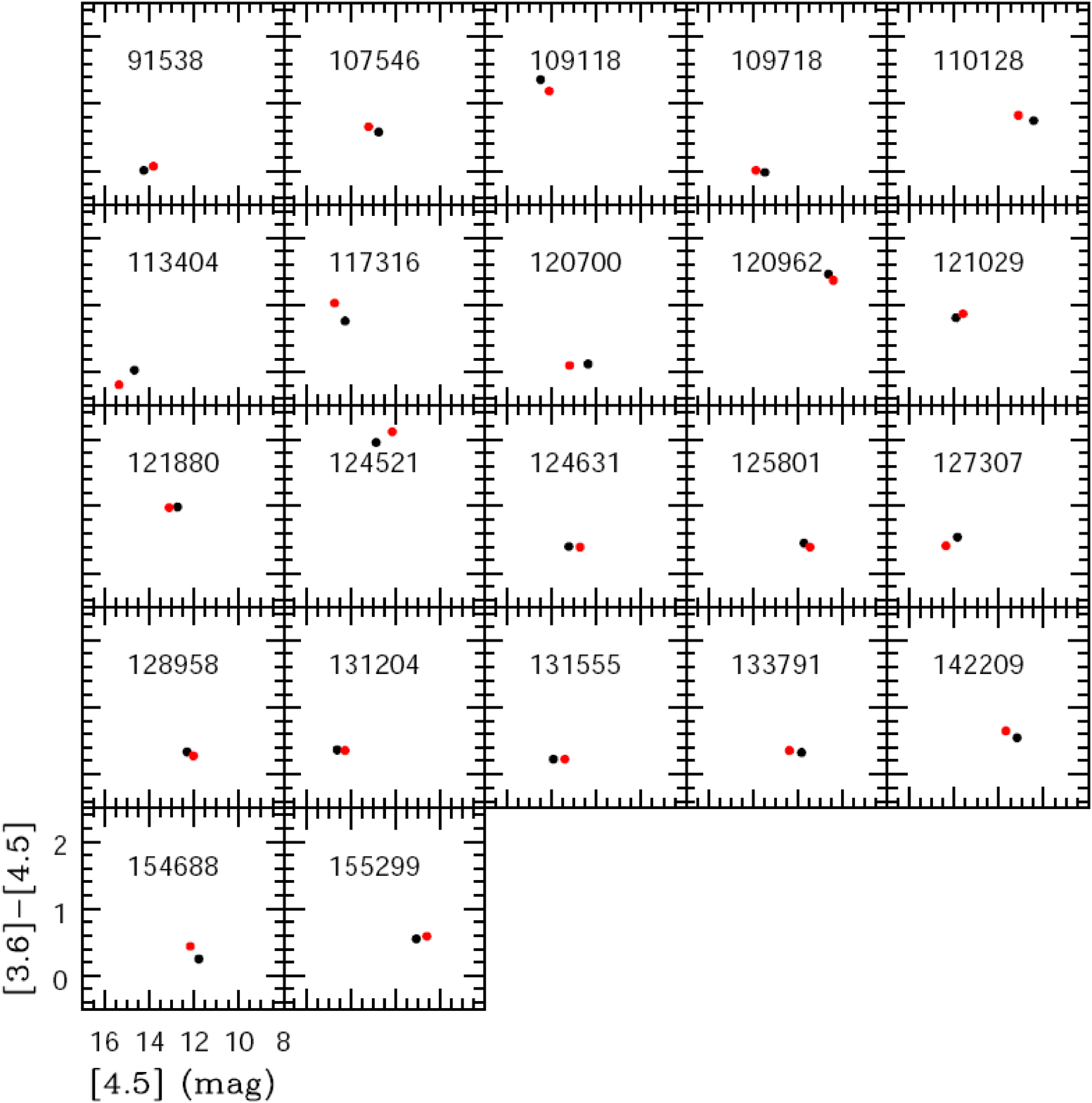}
   \caption{b). As Figure~\ref{colmag2:fig}a, continued.
}
\end{figure}

\begin{figure}
   \centering
   \includegraphics [width=15 cm] {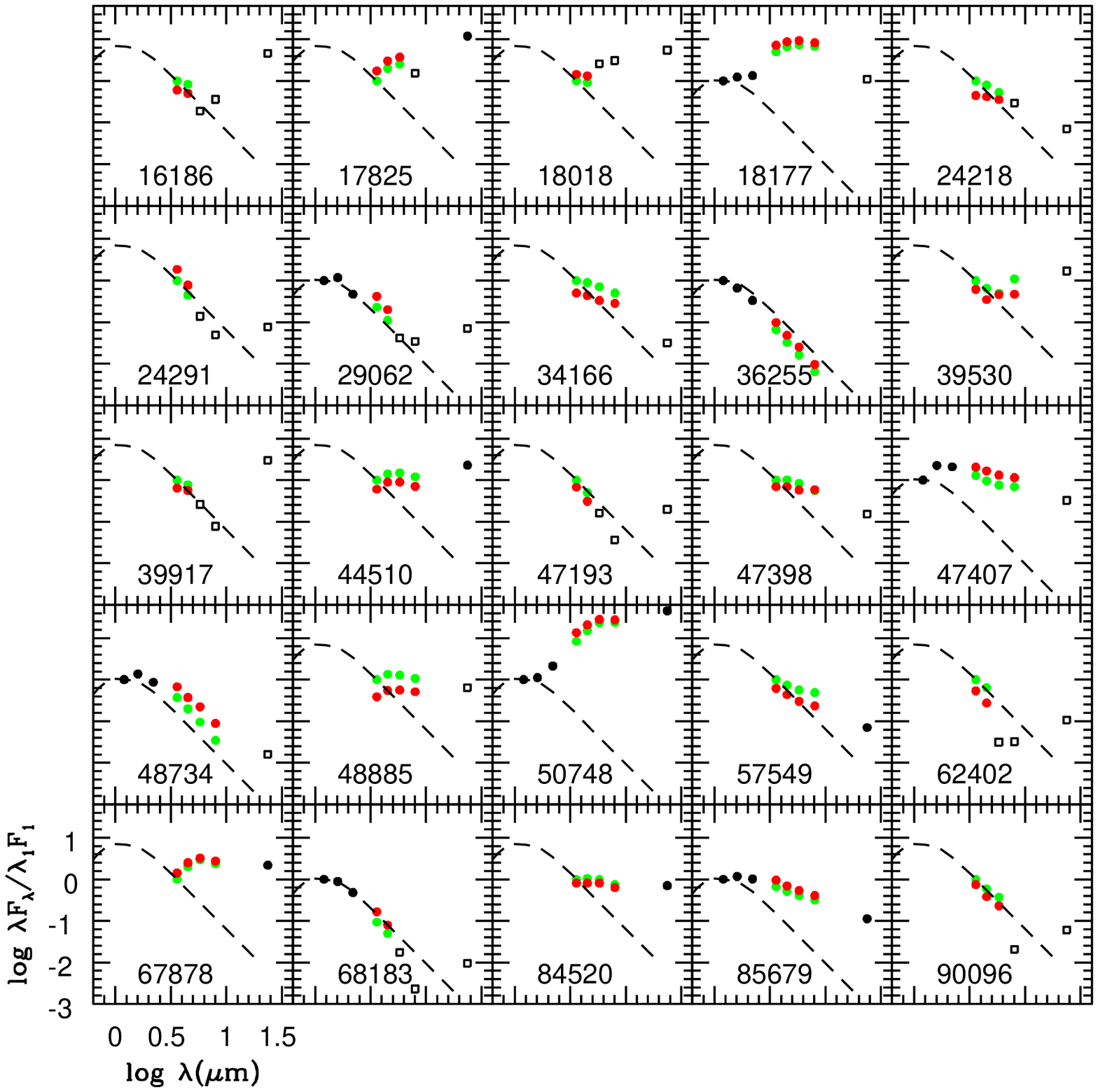}
   \caption{a). SED's of sources as numbered in Table~\ref{mag:tab}. IRAC fluxes observed in the
   first (second) epoch are depicted as green (red) dots. When available, 2MASS fluxes are
   also plotted for completeness although obtained about seven years
   prior to the {\it Spitzer} advent. MIPS fluxes (at 24\,$\mu$m) are also not strictly
   simultaneous to IRAC ones. 2MASS and MIPS 24\,$\mu$m fluxes are plotted as black dots, while
   open squares indicate 3\,$\sigma$ upper limits. Dashed line represents a median
   photosphere of stars in the spectral range K5-M5. As indicated
   by the ordinate label, all the data are normalized to the flux
   corresponding to the shortest available wavelength, namely J or
   3.6\,$\mu$m (this latter for the sources without a 2MASS
   counterpart).
   \label{sedA:fig}}
\end{figure}

\addtocounter{figure}{-1}
\begin{figure}
   \centering
   \includegraphics [width=15 cm] {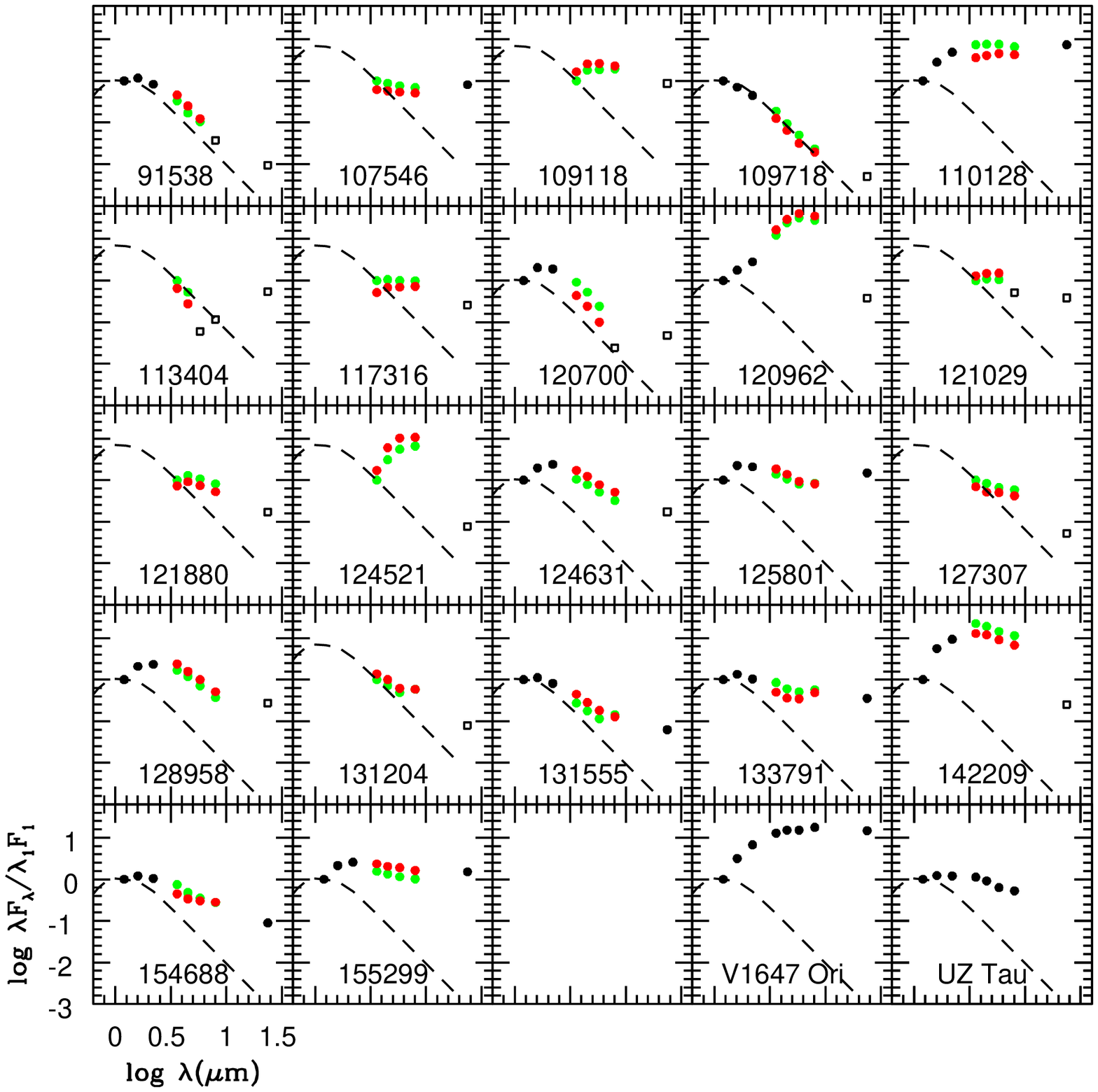}
   \caption{b). As in Figure~\ref{sedA:fig}a. SED's of V1647~Ori and UZ~Tau are shown in the
   last two panels for comparative purposes.
   }
\end{figure}

\newpage

\begin{deluxetable}{cccccc}
\tabletypesize{\normalsize} \tablewidth{0pt}
\tablecaption{Statistics of IRAC variables.\label{stat:tab}}
\tablehead{ Band &   N$_{ep}$$^a$& N$_{m}$$^b$   & N$_{var}$$^c$ & N$_{dec}$$^d$ & N$_{ris}$$^e$ } 
\startdata
\cline{1-6}
3.6 $\mu$m        & 138601   &   40382    &  306 & 190	  & 116\\
4.5 $\mu$m        & 126901   &   27519    &  201 & 108	  & 93 \\
5.8 $\mu$m        & 24256    &    6751    &  121 & 65	  & 56 \\
8.0 $\mu$m        & 12977    &    3207    &  84	 & 45     & 39 \\
3.6 and 4.5 $\mu$m & 114515   &   24482    &  53	 &  -     & -  \\
\cline{1-6}
\enddata
\medskip
\begin{itemize}
\item[a -] Number of sources detected in the first epoch (this number does not change 
significantly in the second epoch).
\item[b -] Number of sources whose magnitude is less than a given threshold (16, 15.5, 13.5
and 12.5 mag in band 1,2,3, and 4, respectively).
\item[c -] Number of sources selected as variables (see text)
\item[d -] Number of declining sources (mag$_{ep1}$-mag$_{ep2}$ $<$ 0)
\item[e -] Number of rising sources (mag$_{ep1}$-mag$_{ep2}$ $>$ 0)
\end{itemize}
\end{deluxetable}

\begin{deluxetable}{ccccccccccc}
\tabletypesize{\scriptsize} \tablewidth{0pt}
\tablecaption{Magnitudes of the selected
variables.\label{mag:tab}} \tablehead{ ident.
&$\alpha$(2000)&$\delta$(2000)&    J    &     H    &    K    &
[3.6]  &[4.5]   &   [5.8]  &  [8.0] &  [24] }
\startdata
16186  & 8 45 29.7 & -43  50  48.1  &   -    &    -     &     -    &  13.50  &  12.96   &     -    &    -   & - \\
       &           &                &   -    &    -     &     -    &  14.04  &  13.51   &     -    &    -   & - \\
 17825  & 8 45 33.9 & -43  52   3.0  &   -    &    -     &     -    &  11.91  &  10.44   &    9.43  &    -  & - \\
        &           &                &   -    &    -     &     -    &  11.32  &   9.99   &    9.01  &    -   & - \\
 18018  & 8 45 34.4 & -43  50   3.8  &   -    &    -     &     -    &  13.12  &  12.49   &  -   &    -    & - \\
        &           &                &   -    &    -     &     -    &  12.74  &  12.10   &  -   &    -   & - \\
 18177  & 8 45 34.8 & -43  52  57.6  &  16.69 &  15.71   &  14.77   &  11.93  &  10.91   &   10.02  &   9.14 & -  \\
        &           &                &   -    &    -     &     -    &  11.53  &  10.60   &    9.76  &   8.91 & - \\
 24218  & 8 45 49.8 & -43  56  56.5  &   -    &    -     &     -    &  14.96  &  14.46   &   14.15  &   -    & - \\
        &           &                &   -    &    -     &     -    &  15.82  &  15.16   &   14.58  &   -    & - \\
 24291  & 8 45 50.0 & -43  56  50.6  &   -    &    -     &     -    &  15.03  &  15.16   &     -    &   -    & - \\
        &           &                &   -    &    -     &     -    &  14.36  &  14.58   &     -    &   -    & - \\
 29062  & 8 46 01.5 & -43  41  20.2  &  16.37 &  15.43   &  15.60   &  14.95  &  15.02   &     -    &  -     & - \\
        &           &                &   -    &    -     &     -    &  14.55  &  14.62   &     -    &  -     & - \\
 34166  & 8 46 14.1 & -43  23  48.0  &   -    &    -     &     -    &  13.11  &  12.52   &   12.02  &  11.39  & - \\
        &           &                &   -    &    -     &     -    &  13.87  &  13.31   &   12.83  &  12.04 & - \\
 36255  & 8 46 19.2 & -43  51   9.4  &  11.52 &  11.22   &  11.13   &  11.46  &  11.49   &   11.48  &  11.49 & - \\
        &           &                &    -   &    -     &  -   &  11.02  &  11.06   &   11.02  &  11.09 & - \\
 39530  & 8 46 27.1 & -43  54  31.2  &    -   &    -     &     -    &  14.68  &  14.43   &   13.97  &  12.10 & - \\
        &           &                &    -   &    -     &     -    &  15.20  &  15.09   &   14.05  &  13.03 & - \\
 39917  & 8 46 28.1 & -43  54  31.9  &    -   &    -     &  -   &  14.96  &  14.52   &  -   &    -   & - \\
        &           &                &    -   &    -     &  -   &  15.46  &  14.86   &  -   &    -   & - \\
 44510  & 8 46 39.5 & -43  22  50.5  &    -   &    -     &  -   &  13.24  &  12.16   &   11.33  &  10.58 &  6.31      \\
        &           &                &    -   &    -     &  -   &  13.80  &  12.64   &   11.89  &  11.14 &
          -        \\
 47193  & 8 46 46.2 & -43  50  24.5  &    -   &    -     &  -   &  14.46  &  14.47   &  -   &    -   &   -        \\
        &           &                &    -   &    -     &  -   &  14.86  &  14.49   &  -   &    -   &   -        \\
 47398  & 8 46 46.7 & -43  19  59.3  &    -   &    -     &  -   &  14.57  &  13.81   &   13.28  &  12.74 &   -        \\
        &           &                &    -   &    -     &  -   &  14.97  &  14.24   &   13.68  &  12.67 &   -        \\
 47407  & 8 46 46.7 & -43  52  52.4  &  16.68 &  15.08   &  14.30   &  13.41  &  13.00   &   12.49  &  11.60 &   -        \\
        &           &                &    -   &    -     &  -   &  12.91  &  12.39   &   11.89  &  11.06 &   -        \\
 48734  & 8 46 49.9 & -43  19   1.2  &  15.69 &  14.59   &  14.26   &  13.75  &  13.71   &   13.74  &  13.86 &   -        \\
        &           &                &    -   &    -     &  -   &  13.10  &  13.03   &   12.82  &  12.83 &   -        \\
 48885  & 8 46 50.3 & -43  52  57.0  &    -   &    -     &  -   &  14.73  &  13.68   &   12.96  &  12.17 &   -        \\
        &           &                &    -   &    -     &  -   &  15.74  &  14.65   &   13.87  &  12.98 &   -        \\
 50748  & 8 46 54.7 & -43  52  36.3  &  17.37 &  16.50   &  14.97   &  12.04  &  10.68   &    9.45  &   8.42 &  4.16      \\
        &           &                &    -   &    -     &  -   &  11.52  &  10.32   &    9.24  &   8.27 &   -        \\
 57549  & 8 47  9.8 & -43  56  31.7  &    -   &    -     &  -   &  13.06  &  12.64   &   12.17  &  11.36 &  9.88      \\
        &           &                &    -   &    -     &  -   &  13.58  &  13.22   &   12.87  &  12.16 &   -        \\
 62402  & 8 47 19.8 & -43  21  20.4  &   -    &    -     &     -    &  14.79  &  14.54   &     -    &   -    &   -        \\
        &           &                &   -    &    -     &     -    &  15.47  &  15.46   &     -    &   -    &   -        \\
 67878  & 8 47 30.6 & -43  28  22.4  &    -   &    -     &  -   &  13.14  &  11.68   &   10.48  &   9.74 &  6.25      \\
        &           &                &    -   &    -     &  -   &  12.77  &  11.43   &   10.38  &   9.58 &   -        \\
 68183  & 8 47 31.3 & -43  55   3.1  &  15.37 &  14.76   &  14.61   &  14.96  &  14.90   &  -   &    -   &   -        \\
        &           &                &    -   &    -     &  -   &  14.33  &  14.40   &  -   &    -   &   -        \\
 84520  & 8 48  0.4 & -43  20  37.2  &    -   &    -     &  -   &  13.36  &  12.59   &   11.91  & 11.22  &  7.70      \\
        &           &                &    -   &    -     &  -   &  13.61  &  12.86   &   12.12  & 11.40  &   -        \\
 85679  & 8 48  2.3 & -43  18  50.5  &  14.34 &  13.42   &  12.73   &  11.77  &  11.33   &   10.80  &  10.09 &  7.66      \\
        &           &                &    -   &    -     &  -   &  11.37  &  10.99   &   10.50  &   9.84 &   -        \\
 90096  & 8 48 10.0 & -43  19  57.9  &    -   &    -     &  -   &  14.39  &  14.23   &   13.96  &    -   &   -        \\
        &           &                &   -    &    -     &  -   &  14.70  &  14.70   &   14.48  &    -   &   -        \\
 91538  & 8 48 12.6 & -43  59  17.4  &  16.05 &  15.15   &  14.69   &  14.26  &  14.25   &   14.01  &    -   &   -        \\
        &           &                &    -   &    -     &  -   &  13.90  &  13.83   &   13.83  &    -   &   -        \\
107546  & 8 48 42.8 & -43  17  32.1  &    -   &    -     &  -   &  13.32  &  12.74   &   12.13  &  11.25 &  7.52      \\
        &           &                &    -   &    -     &  -   &  13.86  &  13.20   &   12.50  &  11.57 &   -        \\
109118  & 8 48 46.0 & -43  37  27.0  &    -   &    -     &  -   &  15.87  &  14.51   &   13.71  &  12.69 &   -        \\
        &           &                &    -   &    -     &  -   &  15.31  &  14.12   &   13.33  &  12.50 &   -        \\
109718  & 8 48 47.3 & -43  45  57.4  &  14.64 &  14.26   &  13.94   &  13.45  &  13.47   &   13.39  &  13.26 &   -        \\
        &           &                &    -   &    -     &  -   &  13.88  &  13.87   &   13.89  &  13.43 &   -        \\
110128  & 8 48 48.1 & -43  56  18.1  &  16.39 &  14.52   &  13.08   &  11.20  &  10.45   &    9.69  &   8.86 &  5.15      \\
        &           &                &    -   &    -     &  -   &  11.96  &  11.13   &   10.23  &   9.32 &   -        \\
113404  & 8 48  0.4 & -43  20  37.2  &   -    &    -     &     -    &  14.70  &  14.67   &     -    &   -    &   -        \\
        &           &                &   -    &    -     &     -    &  15.17  &  15.36   &     -    &   -    &   -        \\
117316  & 8 49  4.1 & -43  20   2.4  &    -   &    -     &  -   &  15.01  &  14.25   &   13.55  &  12.58 &   -        \\
        &           &                &    -   &    -     &  -   &  15.75  &  14.72   &   13.94  &  12.91 &   -        \\
120700  & 8 49 12.4 & -43  17  14.4  &  15.41 &  13.89   &  13.13   &  12.50  &  12.38   &   12.46  &    -   &   -        \\
        &           &                &    -   &    -     &  -   &  13.31  &  13.21   &   13.40  &    -   &   -        \\
120962  & 8 49 13.0 & -43  35  48.1  &  17.83 &  16.47   &  15.12   &  12.09  &  10.63   &    9.56  &   8.73 &   -        \\
        &           &                &    -   &    -     &  -   &  11.78  &  10.41   &    9.32  &   8.49 &   -        \\
121029  & 8 49 13.2 & -43  36  38.4  &    -   &    -     &  -   &  14.74  &  13.93   &   13.21  &    -   &   -        \\
        &           &                &    -   &    -     &  -   &  14.48  &  13.61   &   12.82  &    -   &   -        \\
121880  & 8 49 15.2 & -43  36  33.2  &    -   &    -     &  -   &  13.74  &  12.75   &   12.18  &  11.51 &   -        \\
        &           &                &    -   &    -     &  -   &  14.10  &  13.12   &   12.57  &  11.97 &   -        \\
124521  & 8 49 21.8 & -44   2   1.3  &    -   &    -     &  -   &  14.81  &  12.86   &   11.47  &  10.31 &   -        \\
        &           &                &    -   &    -     &  -   &  14.25  &  12.14   &   10.81  &   9.78 &   -        \\
124631  & 8 49 22.1 & -43  35  42.5  &  16.70 &  15.24   &  14.19   &  13.64  &  13.24   &   12.94  &  12.46 &   -        \\
        &           &                &    -   &    -     &  -   &  13.13  &  12.74   &   12.49  &  11.96 &   -        \\
125801  & 8 49 25.0 & -43  36  12.5  &  15.53 &  13.93   &  13.17   &  12.18  &  11.73   &   11.29  &  10.26 &  6.06      \\
        &           &                &    -   &    -     &  -   &  11.84  &  11.45   &   11.11  &  10.27 &   -        \\
127307  & 8 49 28.7 & -43  38  29.1  &    -   &    -     &  -   &  14.39  &  13.85   &   13.32  &  12.49 &   -        \\
        &           &                &    -   &    -     &  -   &  14.78  &  14.37   &   13.63  &  12.85 &   -        \\
128958  & 8 49 32.7 & -43  15  14.2  &  16.25 &  14.71   &  13.76   &  12.66  &  12.32   &   12.13  &  11.86 &   -        \\
        &           &                &    -   &    -     &  -   &  12.31  &  12.03   &   11.75  &  11.51 &   -        \\
131204  & 8 49 38.2 & -43  51  43.0  &    -   &    -     &  -   &  14.97  &  14.60   &   14.27  &  13.09 &   -        \\
        &           &                &    -   &    -     &  -   &  14.61  &  14.25   &   14.00  &  13.08 &   -        \\
131555  & 8 49 39.1 & -43  39   5.3  &  15.78 &  14.91   &  14.43   &  14.16  &  13.93   &   13.64  &  12.41 &  9.74      \\
        &           &                &    -   &    -     &  -   &  13.65  &  13.42   &   13.15  &  12.53 &   -        \\
133791  & 8 49 44.5 & -43  34  48.8  &  15.01 &  13.93   &  13.38   &  12.16  &  11.83   &   11.23  &  10.14 &  7.09      \\
        &           &                &    -   &    -     &  -   &  12.74  &  12.38   &   11.66  &  10.30 &   -        \\
142209  & 8 50  4.6 & -43  23  51.3  &  18.13 &  15.52   &  14.13   &  11.74  &  11.19   &   10.73  &  10.01 &   -        \\
        &           &                &    -   &    -     &  -   &  12.34  &  11.69   &   11.23  &  10.59 &   -        \\
154688  & 8 50 34.1 & -43  52   1.1  &  14.74 &  13.78   &  13.12   &  12.04  &  11.79   &   11.36  &  10.65 &  8.32      \\
        &           &                &    -   &    -     &  -   &  12.61  &  12.17   &   11.55  &  10.62 &   -        \\
155299  & 8 50 35.5 & -43  33  35.9  &  15.12 &  13.55   &  12.51   &  11.61  &  11.06   &   10.46  &   9.60 &  5.61      \\
        &           &                &    -   &    -     &  -   &  11.18  &  10.59   &    9.92  &   9.10 &   -        \\
\cline{1-11} \cline{1-4}
\enddata
\medskip
\footnotesize
%
\begin{itemize}
\item[-] Source identification refers to our internal numbering (Strafella et al. 2009).
\item[-] All the values are given in magnitudes. Conversion into
flux densities (Jansky) can be done by means of the zero-mag
fluxes that are: 280.9 Jy (3.6\,$\mu$m), 179.7 Jy (4.5\,$\mu$m), 115.0
Jy (5.8\,$\mu$m), 64.13 Jy (8.0\,$\mu$m), and 7.17 Jy (24\,$\mu$m).
\item[-] Typical errors (in mag) are: 2MASS, J: 0.02-0.18, H: 0.02-0.17, K: 0.02-0.15;
IRAC, all bands: 0.1; MIPS\,24\,$\mu$m: 0.3.
\item[-]Magnitudes in other bands:
	\begin{itemize}
	\item[-] 36255: B=14.0, V=13.2, R=12.8
	\item[-] 47193: B=18.8, R=17.6
  	\item[-] 68183: B=13.3, V=17.7, R=17.6
	\item[-] 85675: B=19.5, R=18.2
	\item[-] 91538: R:19.9, I=18.2
	\item[-] 109718: B=18.3, V=16.7, R=16.1
	\item[-] 121029: MIPS70=-1.7
	\item[-] 124521: MIPS70=0.43
	\item[-] 131555: B=21.2, R=20.0
	\item[-] 133791: B=20.9, R=19.0
	\item[-] 154688: B=19.5, R=17.7
	\item[-] 155299: R(phot F)=19.5
 	\end{itemize}
\end{itemize}
\end{deluxetable}

\begin{deluxetable}{lcccccc}
\tabletypesize{\footnotesize} \tablecaption{Coded photometric
properties of the selected sources. \label{summary:tab}}
\tablewidth{0pt} \tablehead{
\colhead{Id.$^a$} & \colhead{location$^b$} &  \colhead{YSO vs.} &\colhead{$\Delta$mag$^c$}&
\colhead{[3.6]-[4.5]$^d$} & $\mathcal{E}$$^e$& L$_{IRAC}$ \\
        &       &      other        & \colhead{vs. $\lambda$}   & \colhead{vs. [4.5]}& &(L$_{\sun}$)  }
\startdata
%
16186         & IRS16$^f$  & YSO     &  C   & 0 & 0.7 &  0.004 \\ 
{\bf 17825}   & IRS16$^f$  & YSO     &  D   & - & 4.1 &  0.101 \\
{\bf 18018}   & IRS16$^f$  & YSO     &  C   & 0 & 1.2 &  0.008 \\
{\bf 18177}   & IRS16$^f$  & YSO     &  D   & - &20.0 &  0.072 \\
24218         &   I        & AGB/XGAL&  D   & - & 1.0 &  0.001 \\
24291         &   I        & PHT     &  D   & - & 1.0 &  0.001 \\
29062         &   O        & XGAL    &  C   & 0 & 1.7 &  0.001 \\
34166         &   O        & YSO     &  C   & 0 & 1.5 &  0.009 \\
36255         &   I        & PHT/AGB &  C   & 0 & 1.0 &  0.037 \\
39530         &   P        & XGAL    &  I   & + & 2.0 &  0.002 \\
39917         &   P        & PHT     &  D   & - & 1.0 &  0.001 \\
{\bf 44510}   &   I        & YSO     &  C   & - & 3.0 &  0.015 \\
47193         &   I        & PHT     &  I   & + & 1.0 &  0.001 \\
{\bf 47398}   &   I        & YSO     &  C   & 0 & 1.5 &  0.003 \\
47407         &   P        & YSO     &  C   & + & 4.6 &  0.011 \\
48734         &   I        & PHT     &  I   & 0 & 2.4 &  0.006 \\
{\bf 48885}   &   P        & YSO     &  D   & - & 3.0 &  0.003 \\
{\bf 50748}   &   P        & YSO     &  D   & - &56.0 &  0.107 \\
57549         &   O        & AGB     &  I   & + & 1.5 &  0.009 \\
62402         &   I        & PHT     &  I   & + & 1.0 &  0.001 \\
{\bf 67878}   &   P        & YSO     &  D   & - & 6.3 &  0.038 \\
68183         &   I        & PHT     &  D   & - & 1.2 &  0.001 \\
{\bf 84520}   &   I        & YSO     &  C   & 0 & 2.2 &  0.011 \\
85679         &   I        & AGB     &  D   & - & 2.6 &  0.045 \\
90096         &   I        & PHT     &  I   & + & 1.0 &  0.002 \\
91538         &   O        & PHT/AGB &  D   & + & 1.7 &  0.003 \\
{\bf 107546}  &   P        & YSO     &  D   & - & 1.8 &  0.009 \\
{\bf 109118}  &   I        & YSO     &  D   & - & 6.6 &  0.003 \\
109718        &   I        & PHT/AGB &  D   & 0 & 1.2 &  0.004 \\
{\bf 110128}  &   I        & YSO     &  D   & - &16.0 &  0.072 \\
113404        &   I        & XGAL    &  I   & + & 1.0 &  0.001 \\
117316        &   P        & XGAL    &  D   & - & 2.0 &  0.002 \\
120700        &   P        & PHT     &  I   & 0 & 2.9 &  0.009 \\
{\bf 120962}  &   P        & YSO     &  C   & - &50.0 &  0.096 \\
121029        &   P        & YSO     &  I   & + & 3.0 &  0.004 \\
{\bf 121880}  &   P        & YSO     &  C   & 0 & 2.7 &  0.008 \\
124521        &   I        & YSO     &  C   & + &20.0 &  0.020 \\
{\bf 124631}  &   P        & YSO     &  C   & 0 & 5.0 &  0.008 \\
{\bf 125801}  &   I        & YSO     &  D   & - & 4.6 &  0.029 \\
127307        &   O        & YSO     &  C   & + & 1.5 &  0.003 \\
128958        &   I        & PHT     &  C   & - & 4.8 &  0.016 \\
131204        &   O        & XGAL    &  D   & 0 & 2.0 &  0.002 \\
{\bf 131555}  &   I        & YSO     &  D   & 0 & 2.0 &  0.004 \\
{\bf 133791}  &   I        & YSO     &  D   & 0 & 2.7 &  0.022 \\
{\bf 142209}  &   P        & YSO     &  C   & - &42.0 &  0.033 \\
154688        &   O        & AGB     &  D   & - & 2.4 &  0.022 \\
155299        &   O        & YSO     &  C   & 0 & 5.9 &  0.064 \\
\hline
EXor class    & P(I)       & YSO     &D(C)  &-(0)& $>$ 1.0 &        \\
\enddata
\tablenotetext{a}{~~ In bold-face the candidate young variables are indicated}
\tablenotetext{b}{~~ O = outside the CO contours, I = inside, P = on a peak of warm gas.}
\tablenotetext{c}{~~ D = decreasing, C = constant, I = increasing.}
\tablenotetext{d}{~~ - = redder, 0 = constant, + = bluer color during a fading event.}
\tablenotetext{e}{~~ ratio of the observed SED and the underlying normalized K5-M5 photosphere, both integrated
from $J$ to 8\,$\mu$m.}
\tablenotetext{f}{~~~~ denotes that the source lies within the IR cluster IRS16, although located outside our CO map.}

\end{deluxetable}


\begin{thebibliography}{}

\bibitem{} Cutri, R.M., Strutskie, M.F., Van Dyk, S. et al. 2003 Explanatory
Supplement to the 2MASS All Sky Data Release (Pasadena: Caltech)
\bibitem{} De Luca, M., Giannini, T., Lorenzetti, D. et al. 2007 A\&A 474, 863
\bibitem{} D'Alessio, P., Calvet, N., Hartmann, L., Lizano, S. \& Cant\'{o}, J. 1999 ApJ 527, 893
\bibitem{} Elia, D., Massi, F., Strafella, F. et al. 2007 ApJ 655, 316
\bibitem{} Fazio, G.G., Hora, J.L., Allen, L.E. et al. 2004 ApJS 154, 10
\bibitem{} J{\o}rgensen, J.K., Harvey, P.M., Evans II, N.J., et al. 2006 ApJ 645, 1246
\bibitem{} Giannini, T., Lorenzetti, D., De Luca, M. et al. 2007 ApJ 671, 470
\bibitem{} Giannini, T., Massi, F., Podio L. et al. 2005 A\&A 433, 941
\bibitem{} Greene, T.P., Wilking, B.A., Andr\`e, P., Young, E.T., Lada, C.J. 1994 ApJ 434,614
\bibitem{} Hartmann, L. \& Kenyon, S. 1985 ApJ 299, 462
\bibitem{} Hartmann, L. \& Kenyon, S. 1996 ARAA 34, 207
\bibitem{} Hartmann, L., Megeath, S.T., Allen, L. et al. 2005 ApJ 629, 881
\bibitem{} Harvey, P.M., Chapman, N., Shih-Ping, L. et al. 2006 ApJ 644, 307
\bibitem{} Herbig, G.H. 1989 Proc. of the ESO Workshop on {\it Low Mass Star Formation and
Pre-Main Sequence Objects}, ed. B. Reipurth, p.233
\bibitem{} Herbig, G.H. 2007 AJ 133, 2679
\bibitem{} Herbig, G.H. 2008 AJ 135, 637
\bibitem{} Hern\'{a}ndez, J., Hartmann, L., Megeath, T. et al. 2007 ApJ 662, 1067
\bibitem{} Hodapp, K.W. 1999 AJ 118, 1338
\bibitem{} Hodapp, K.W., Hora, J.L., Rayner, J.T. et al. 1996 ApJ 468, 861
\bibitem{} K\'{o}sp\'{a}l, \'{A}., \'{A}brah\'{a}m, P. , Acosta-Pulido, J. et al. 2007 A\&A 470, 211
\bibitem{} Lorenzetti, D., Giannini, T., Calzoletti, L. et al. 2006 A\&A 453, 579 
\bibitem{} Lorenzetti, D., Giannini, T., Larionov, V.M. et al. 2007 ApJ 665, 1182 (L07)
\bibitem{} Lorenzetti, D., Giannini, T., Vitali, F. et al. 2002 ApJ 564, 839
\bibitem{} Lorenzetti, D., Larionov, V.M., Giannini, T. et al. 2009, ApJ 693, 1056 (L09)
\bibitem{} Marengo, M., Reiter, M. \& Fazio, G.G. 2008, in AIP Conf. Proc. 1001,
IXth Torino Workshop on Evolution and Nucleosynthesis in AGB stars and the 2nd Perugia Workshop on
Nuclear Astrophysics. (New York, AIP), 331
\bibitem{} Massi, F., De Luca, M., Elia, D. et al. 2007 A\&A 466, 1013
\bibitem{} Massi, F., Lorenzetti, D. \& Giannini, T. 2003 A\&A 399, 147
\bibitem{} Massi, F., Lorenzetti, D., Giannini, T. \& Vitali, F. 2000 A\&A 353, 598
\bibitem{} Massi, F., Giannini, T., Lorenzetti, D. et al. 1999 A\&AS 136, 471
\bibitem{} Muzerolle, J., Megeath, S.T., Flaherty, K.M. et al. 2005 ApJ 620, 107
\bibitem{} Porras, A., J{\o}rgensen, J.K., Allen, L.E., et al. 2007 ApJ 656, 493
\bibitem{} Price, S.D., Egan, M.P., Carey, S.J., Mizuno, D.R., \& Kuchar, T.A. 2001 AJ 121, 2819
\bibitem{} Rebull, L.M., Stapelfeldt, K.R., Evans II, N.J et al. 2007 ApJ, 171, 447 
\bibitem{} Reipurth, B. \& Aspin, C. 2004 ApJ 606, L119
\bibitem{} Riaz, B, Muller, D.J., \& Grizis, J.E. 2006 ApJ, 650, 1133 
\bibitem{} Rieke, G.H., Young, E.T., Engelbracht, C.W. et al. 2004 ApJS 154, 24
\bibitem{} Robitaille, T.P., Meade, M.R., Babler, B.L. et al. 2008 AJ 136, 2413
\bibitem{} Rowan-Robinson, M., Lari, C., Perez-Fournon, I., et al. 2004 MNRAS 351, 1290
\bibitem{} Sicilia-Aguilar, A., Mer\'{\i}n, B., T., Hormuth, F. \& \'{A}brah\'{a}m, P. 2008 ApJ 673, 382
\bibitem{} Schuster, M. T., Marengo, M., Patten, B. M. 2006 SPIE meeting,  
Orlando, FL, \#6720-65
\bibitem{} Strafella, F., et al. 2009 ApJ, in preparation
\bibitem{} Shu, F.H., Najita, J.R., Ostriker, E. et al. 1994 ApJ 429, 781
\bibitem{} Stetson, P.B. 1987 PASP 99, 191
\bibitem{} Werner, M.W., Roelling, T.L., Low, F.J. et al. 2004 ApJS 154, 1
\bibitem{} Winston, E., Megeath, S.T., Wolk, S.J. et al. 2007 ApJ 669, 493
\bibitem{} Winston, E., Megeath, S.T., Wolk, S.J. et al. 2009 ApJ, in press (arXiv0904.1244W)

\end{thebibliography}
\end{document}